\documentclass[fleqn]{2023SCGE}
\usepackage{color}
\usepackage{float}
\usepackage{rotfloat}
\usepackage{rotating}
\usepackage{epsfig}
\usepackage{booktabs}
\usepackage{amsmath}
\usepackage{graphicx}
\usepackage{natbib,times}
\usepackage{amsmath}
\usepackage{url}
\usepackage{hyperref}
\usepackage{pdflscape}
\setcitestyle{numbers,square}

\begin{document}

\ensubject{subject}

\ArticleType{Article}
\Year{2023}
\Month{??}
\Vol{??}
\No{??}
\DOI{10.1007/??}
\ArtNo{??}
\ReceiveDate{?? ??, 2023}
\AcceptDate{?? ??, 2023}

\title{The Nearest Neutron Star Candidate in a Binary Revealed by Optical Time-domain Surveys}

\author[1,$\dagger$]{Ling-Lin Zheng}{}

\author[1,$\dagger$]{Mouyuan Sun}{}

\author[1]{Wei-Min Gu}{{guwm@xmu.edu.cn}}

\author[1]{Tuan Yi}{}

\author[1]{Zhi-Xiang Zhang}{{zhangzx@xmu.edu.cn}}

\author[2,3]{Pei Wang}{}

\author[1]{\\Junfeng Wang}{}

\author[1]{Jianfeng Wu}{}

\author[4]{Shan-Shan Weng}{}

\author[2]{Song Wang}{}

\author[1]{Sen-Yu Qi}{}

\author[5]{Jia Zhang}{}

\author[2]{\\Chun-Qian Li}{}

\author[2,6]{Jian-Rong Shi}{}

\author[7,8]{Yong Shao}{}

\author[7,8]{Xiang-Dong Li}{}

\author[1]{Jin-Bo Fu}{}

\author[2]{Fan Yang}{}

\author[2]{Zhongrui Bai}{}

\author[2]{\\Yu Bai}{}

\author[2]{Haotong Zhang}{}

\author[2,6]{Jifeng Liu}{{jfliu@nao.cas.cn}}

\AuthorMark{Zheng L L}

\AuthorCitation{Zheng L L, Sun M Y, Gu W M, et al}
\footnotetext[1]{$\dagger$ Ling-Lin Zheng and Mouyuan Sun contributed equally to this work.}
\footnotetext[2]{$*$ Corresponding author(s). E-mail(s): guwm@xmu.edu.cn; zhangzx@xmu.edu.cn; 
jfliu@nao.cas.cn.}

\address[1]{Department of Astronomy, Xiamen University, Xiamen, 361005, China}
\address[2]{National Astronomical Observatories, Chinese Academy of Sciences, Beijing, 100101, China}
\address[3]{Institute for Frontiers in Astronomy and Astrophysics, Beijing Normal University, Beijing, 102206, China}
\address[4]{Department of Physics and Institute of Theoretical Physics, Nanjing Normal University, Nanjing, 210023, Jiangsu, China}
\address[5]{Yunnan Observatories, Chinese Academy of Sciences, Kunming, 650216, Yunnan, China}
\address[6]{College of Astronomy and Space Science, University of Chinese Academy of Sciences, Beijing, 100049, China}
\address[7]{School of Astronomy and Space Science, Nanjing University, Nanjing, 210023, Jiangsu, China}
\address[8]{Key Laboratory of Modern Astronomy and Astrophysics, Nanjing University, Nanjing, 210023, Jiangsu, China}

\abstract{
	The near-Earth (within $\sim 100$ pc) supernova explosions 
  \citep[e.g.,][]{Ellis1996, Wallner2021} 
	in the past several million years can cause the global deposition of radioactive elements 
	(e.g., $^{60}$Fe) on Earth. The remnants of such supernovae are too old to be easily 
	identified. It is therefore of great interest to search for million-year-old near-Earth 
	neutron stars or black holes, the products of supernovae. However, neutron stars and black 
	holes are 
	challenging to find even in our Solar neighbourhood if they are not radio pulsars 
	or X-ray/$\gamma$-ray emitters. Here we report the discovery of one of the nearest ($127.7 
	\pm 0.3$ pc) neutron star candidates in a detached single-lined spectroscopic binary 
	LAMOST J235456.73+335625.9 (hereafter J2354). Utilizing the time-resolved ground-based 
	spectroscopy and space photometry, we find that J2354 hosts an unseen compact object 
	with $M_{\mathrm{inv}}$ being $1.4 \sim 1.6\ M_{\odot}$. The follow-up \textit{Swift} 
	ultraviolet (UV) and X-ray observations suggest that the UV and X-ray emission is 
	produced by the visible star rather than the compact object. Hence, J2354 
	probably harbours a neutron star rather than a hot ultramassive white dwarf. 
	Two-hour exceptionally sensitive radio follow-up observations with Five-hundred-meter 
	Aperture Spherical radio Telescope fail to reveal any pulsating radio signals at the 
	$6\sigma$ flux upper limit of $12.5\ \mu \mathrm{Jy}$. Therefore, the neutron star 
	candidate in J2354 can only be revealed via our time-resolved observations. 
	Interestingly, the distance between J2354 and our Earth can be as close as 
	$\sim 50$ pc around $2.5$ Myrs ago, as revealed by the Gaia kinematics. Our discovery 
	demonstrates a promising way to unveil the hidden near-Earth neutron stars in binaries 
	by exploring the optical time domain, thereby facilitating understanding of the 
	metal-enrichment history in our Solar neighbourhood.
}



\maketitle

\begin{multicols}{2}
	
\section{Introduction}\label{sect:intro}

As one of the most remarkable explosions, nearby supernovae can have great impacts on 
our Earth and other planets in the Solar system. Heavy elements produced by the supernova 
progenitors spread out because of the powerful supernova shocks and sink to the surfaces 
of Earth and other planets. Indeed, the radioactive elements (e.g., $^{60}$Fe) from 
supernovae in the past million years are detected in Earth's deep sea 
\citep[e.g.,][]{Wallner2021}. Intensive $\gamma$-ray photons or cosmic rays emitted by 
the nearby supernovae may cause disastrous changes to Earth's ecosystem and lead to 
catastrophic extinction events. Hence, the demographics of near-Earth neutron stars and 
black holes, the possible ``fossils'' of aforementioned supernovae, is of great importance 
to various research fields. 

Traditional neutron-star searching methods often aim to detect radio pulsars, X-ray or 
$\gamma$-ray emitters \citep[e.g.,][]{Harding2013}. The majority of inactive neutron 
stars in our Solar neighborhood remain to be discovered. Optical time-domain surveys 
act as a supplementary method for hunting neutron stars in binaries by measuring the 
radial velocities ($V_{\mathrm{r}}$), the masses ($M_{\mathrm{vis}}$) of the visible 
companions, and the orbital periods ($P_{\mathrm{orb}}$). This approach has been 
proven successful in finding several stellar black holes 
\citep[e.g.,][]{Casares2014,Liu2019, Thompson2019, Jayasinghe2021, Shenar2022, 
Chakrabarti2023, El-Badry2023, El-Badry2023-BH1, Tanikawa2023}, 
albeit some candidates might be controversial \citep{Abdul-Masih2020}, and is expected 
to substantially increase the sample size of non-interacting black hole binaries 
\citep{Yi2019, Wiktorowicz2020,Mu2022,Mu2022-update}. The same methodology should 
also be efficient in discovering neutron stars. 

To date, only a few neutron star candidates are discovered \citep[e.g.,][]{Andrews2022, 
Knight2022, Mazeh2022, Yi2022, Yuan2022, Lin2023, Shahaf2023} via time-resolved observations. 
These candidates are far from our Solar system ($\geq 300$ pc). By comparison, 
the nearest pulsar, PSR J0437-4715, locates only $156.8\pm 0.25$ pc away \citep{Reardon2016}. 
The distance of the nearest neutron star, RX J185635-3754 \citep{Walter1996}, which 
is identified by its thermal X-ray emission, is $123^{+11}_{-15}$ pc \citep{Walter2010}. 
Neutron stars in X-ray binaries are often identified via type I X-ray bursts. Among 
them, Cen X-4 is the nearest one with a rather large distance of $1$~kpc \citep{Kuulkers2009}. 
X-ray faint or radio-quiet neutron stars in our Solar neighborhood ($\lesssim 150$ pc) 
remain to be discovered. 

The Large Sky Area Multi-Object Fiber Spectroscopic Telescope (LAMOST) medium-resolution 
(MRS) time-domain survey is conducting repeated 
spectroscopic observations since October 2018 \citep{Liu2020,Zong2020}. The 
spectra from the LAMOST time-domain survey (with a spectral resolution of 
$R\approx7500$) cover the wavelength ranges in [4950 \AA, 5350 \AA] and [6300 \AA, 
6800 \AA] for the blue and red arms \citep{Hou2018, Zhang2020}, respectively. 

We aim to find nearby binaries that harbour unseen neutron stars or black holes by 
exploring the LAMOST medium-resolution time-domain spectroscopic surveys. 
Our selection criteria are as follows. 
\begin{itemize}
  \item with more than three repeated high signal-to-noise (S/N $\geq10$) 
  LAMOST observations. 
  \item are not classified as eclipsing binaries. 
  \item with radial-velocity variations $\Delta V_{\mathrm{r}} \geq 80\ 
  \mathrm{km\ s^{-1}}$ (i.e., like host a massive companion). 
  \item close to our Earth. 
\end{itemize}
Among the $\sim 2\times 10^5$ stars with the LAMOST medium-resolution time-domain 
spectroscopic coverage, J2354 is extraordinary because of its unusually large 
radial-velocity 
($V_{\mathrm{r}}$) variations ($\lvert \Delta V_{\mathrm{r}} \rvert \simeq 400\ 
\mathrm{km\ s^{-1}}$) and small distance \citep{Gaia} to Earth ($127.7\pm 0.3$ pc). 
A systematic study of its archival and follow-up observations suggest that J2354 
hosts a neutron star candidate. If confirmed, the neutron star is the nearest one 
detected in binaries.\footnote{A neutron star candidate at a 
similar distance of J2354 was recently reported by \cite{Lin2023}. Its mass 
is $0.98\pm 0.03\ M_{\odot}$, less massive than the known neutron 
stars \citep[e.g.,][]{Kiziltan2013, Rocha2021}.} The supernova that produces 
the neutron star may significantly alter the Solar environment. 

The manuscript is formatted as follows. In Section~\ref{sect:data_method}, we present 
the multi-band, multi-epoch spectroscopic, and photometric observations of J2354. 
In Section~\ref{sect:result}, we provide doppler spectroscopy evidence that J2354 
harbours a neutron star candidate. In Section~\ref{sect:discussion}, we present our 
efforts to detect radio, X-ray, and $\gamma$-ray signatures of the neutron star 
candidate. Summary and future prospects are made in Section~\ref{sect:sum}.

\section{Data and Methodology}\label{sect:data_method}

\subsection{The Optical Spectroscopic Observations}
\subsubsection{The LAMOST Spectra}\label{sect:lamost}
\begin{figure*}
  \centering
  \includegraphics[keepaspectratio, width=0.5\textwidth]{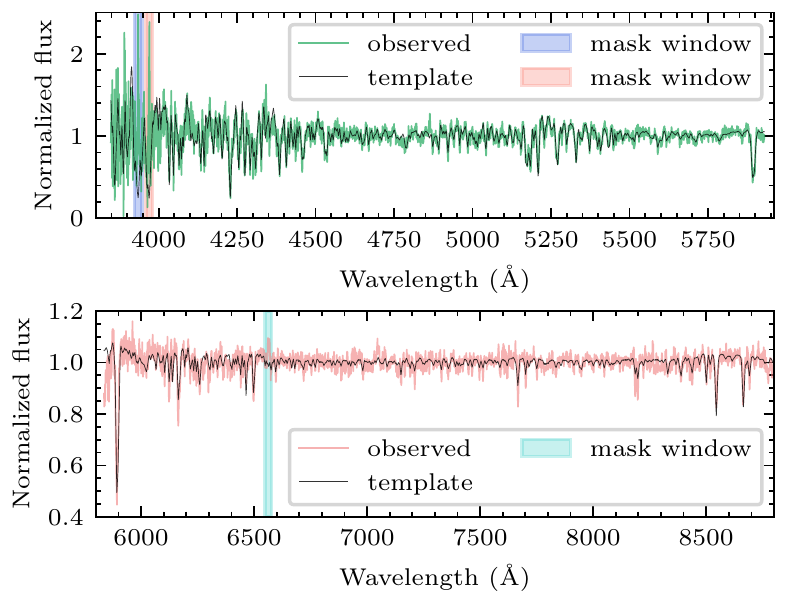} \\
  \caption{\textbf{A LAMOST spectrum for the blue (upper panel) 
  and red arms (lower panel) for J2354.} The best-matching spectral template 
  is shown in black. In the red arm, there is an evident H$\alpha$ emission 
  line, and this line is masked during the radial-velocity measurements. }
  \label{fig:specmatch}
\end{figure*}

J2354 has the optical position of RA=358.736516 deg and DEC=33.940474 deg 
(J2000.0 coordinates) and its LAMOST spectra are consistent with a single-lined 
K7 dwarf star whose V-band magnitude is $13.61\pm 0.02$ mag. The 
single-lined spectroscopic binary (Figure~\ref{fig:specmatch}) has 22 
medium-resolution spectra in LAMOST 
Data Releases 8 and 9 (hereafter LDR8 and LDR9). Additional low-resolution 
(LRS; $R\approx 1800$) spectroscopic observations are obtained from the LAMOST 
Data Release 7 (hereafter LDR7); the spectra cover the wavelength ranges in 
[3690 \AA, 9100 \AA]. 

Absorption lines in J2354 LAMOST spectra show consistent large Doppler shifts; 
in fact, the spectra obtained in the same night can reach a maximum of $\sim 
230\ \mathrm{km\ s^{-1}}$ (see Table~\ref{tbl:rv}). 

\subsubsection{The CFHT Spectra}
\label{sect:cfht}
We requested two CFHT (Canada France Hawaii Telescope) follow-up observations 
on August 16, 2022, to obtain the high-resolution optical spectra of J2354. 
The purpose of the CFHT observations is twofold: to obtain accurate radial 
velocities and to determine the rotational broadening velocity 
$V_{\mathrm{rot}}\sin i$ of the visible star, where $V_{\mathrm{rot}}$ and 
$i$ are the star's rotation velocity and the inclination angle, respectively. 
Hence, we took two on-target 20-minute exposures using the ESPaDOnS instrument 
in the ``object+sky'' spectroscopic mode. The wavelength coverage is from 
$3690$ to $10480$ $\mathrm{\AA}$ and the spectral resolution is $\sim 68000$. 
The two exposures were carefully arranged near the 0.75 quadrature phase of 
the visible star. As a result, during the two exposures, the ``acceleration'' 
of the radial velocity is minimized and only makes negligible contributions 
($\sim 1\ \mathrm{km\ s^{-1}}$) to the observed line broadening. 

The CFHT spectra are reduced with the \texttt{OPERA} 
(\url{https://www.cfht.hawaii.edu/en/projects/opera/index.php}) 
automatic pipeline developed by the CFHT data processing team. Cosmic rays 
contaminate some pixels of the reduced spectra. We use the conventional 
3-sigma clipping to remove cosmic rays. 

\subsubsection{Measuring the Radial Velocities}
The radial velocities of J2354 are measured via the standard cross-correlation 
function (CCF) method. The procedures can be briefly described 
as follows. First, \textit{PyHammer} 
(\href{https://github.com/BU-hammerTeam/PyHammer}{PyHammer link}), 
a Python tool to quickly and automatically classify stellar 
spectra \citep{Kesseli2017, Kesseli2020, Roulston2020} by comparing the observed 
spectra with templates, are used to determine the best-matching spectral type 
of the visible star in J2354 (see Figure~\ref{fig:specmatch}). Second, the 
following wavelength window, [4910 \AA, 5375 \AA], is selected to normalize 
the LAMOST continuum spectra; note that 
cosmic rays are rejected via the median filter method. The best-matching K7 
template is normalized via our \textit{spectool} 
(\url{https://gitee.com/zzxihep/spectool}) package. Third, the standard CCF 
technique is applied to the LAMOST/CFHT spectra and our selected templates 
to measure the radial velocities. The radial-velocity uncertainties are 
estimated via the popular ``flux randomization/random subset sampling 
(FR/RSS)" method \citep{Peterson1998}, which randomizes the observed spectra 
via the following two steps: first, random subsets of the observed spectra 
are selected by adopting the bootstrapping method; second, Gaussian white 
random noise (whose standard deviations are fixed to the observed flux 
uncertainties) are added to the fluxes of the random subsets. 

\begin{table*}[!t]
\begin{center}
  \caption{The LAMOST and CFHT observation log}
  \label{tbl:rv}
  \begin{tabular}{crrrrr}
  \hline
  Spectrograph  & HJD & Phase & RV(absorption lines) & RV(${\mathrm{H\alpha}}$) \\ \hline
             &  day &  &  $\mathrm{km\ s^{-1}}$ & $\mathrm{km\ s^{-1}}$ & \\ \hline
  (1)       & (2) & (3) & (4) & (5) \\ \hline
LAMOST-MRS &	2458822.94999 	&	0.976 	&$	73 	^{+	1.5 	}_{ -	2.2 	}$&$	73 	^{+	2.2 	}_{ -	1.0 	}$	\\
 &	2458822.96622 	&	0.009 	&$	25 	^{+	2.2 	}_{ -	2.2 	}$&$	43 	^{+	6.2 	}_{ -	2.3 	}$	\\
 &	2458822.98244 	&	0.043 	&$	-20 	^{+	2.2 	}_{ -	2.3 	}$&$	-10 	^{+	20.3 	}_{ -	9.1 	}$	\\
 &	2458823.04789 	&	0.180 	&$	-159 	^{+	1.5 	}_{ -	1.4 	}$&$	-158 	^{+	6.2 	}_{ -	6.2 	}$	\\
 &	2458825.93698 	&	0.200 	&$	-168 	^{+	1.0 	}_{ -	1.0 	}$&$	-169 	^{+	6.2 	}_{ -	3.2 	}$	\\
 &	2458825.95118 	&	0.229 	&$	-173 	^{+	1.0 	}_{ -	5.2 	}$&$	-178 	^{+	6.2 	}_{ -	3.2 	}$	\\
 &	2458825.98035 	&	0.290 	&$	-173 	^{+	5.2 	}_{ -	1.0 	}$&$	-165 	^{+	14.3 	}_{ -	7.2 	}$	\\
 &	2458831.94381 	&	0.716 	&$	255 	^{+	1.0 	}_{ -	1.0 	}$&$	236 	^{+	8.1 	}_{ -	5.1 	}$	\\
 &	2458831.96003 	&	0.750 	&$	260 	^{+	1.0 	}_{ -	1.0 	}$&$	250 	^{+	6.2 	}_{ -	6.2 	}$	\\
 &	2458831.97625 	&	0.783 	&$	255 	^{+	1.0 	}_{ -	1.0 	}$&$	256 	^{+	6.2 	}_{ -	6.2 	}$	\\
 &	2458831.99264 	&	0.818 	&$	240 	^{+	1.0 	}_{ -	1.0 	}$&$	237 	^{+	1.0 	}_{ -	1.0 	}$	\\
 &	2458832.00885 	&	0.851 	&$	214 	^{+	5.2 	}_{ -	1.0 	}$&$	213 	^{+	6.2 	}_{ -	1.0 	}$	\\
 &	2459130.11192 	&	0.003 	&$	35 	^{+	2.2 	}_{ -	1.4 	}$&$	43 	^{+	4.3 	}_{ -	4.3 	}$	\\
 &	2459130.12845 	&	0.037 	&$	-14 	^{+	1.4 	}_{ -	1.4 	}$&$	-16 	^{+	6.2 	}_{ -	3.2 	}$	\\
 &	2459130.15552 	&	0.094 	&$	-85 	^{+	1.5 	}_{ -	1.4 	}$&$	-91 	^{+	1.0 	}_{ -	1.0 	}$	\\
 &	2459130.17179 	&	0.128 	&$	-117 	^{+	1.4 	}_{ -	1.4 	}$&$	-122 	^{+	6.2 	}_{ -	6.2 	}$	\\
 &	2459130.19079 	&	0.167 	&$	-151 	^{+	2.3 	}_{ -	1.4 	}$&$	-152 	^{+	1.0 	}_{ -	1.0 	}$	\\
 &	2459130.20705 	&	0.201 	&$	-168 	^{+	1.4 	}_{ -	2.2 	}$&$	-170 	^{+	6.2 	}_{ -	1.0 	}$	\\
 &	2459185.95841 	&	0.369 	&$	-120 	^{+	3.4 	}_{ -	2.2 	}$&$	-122 	^{+	7.2 	}_{ -	4.2 	}$	\\
 &	2459185.97463 	&	0.403 	&$	-83 	^{+	2.2 	}_{ -	2.3 	}$&$	-88 	^{+	9.1 	}_{ -	5.1 	}$	\\
 &	2459185.99087 	&	0.437 	&$	-41 	^{+	3.2 	}_{ -	3.4 	}$&$	-43 	^{+	10.2 	}_{ -	5.3 	}$	\\
 &	2459186.01184 	&	0.481 	&$	15 	^{+	3.2 	}_{ -	3.4 	}$&$	5 	^{+	7.2 	}_{ -	4.2 	}$	\\
  \hline
  LAMOST-LRS &	2458101.95581 	&	0.654 	&$	225 	^{+	3.2 	}_{ -	4.1 	}$&$	206 	^{+	32.7 	}_{ -	15.0 	}$	\\
 &	2458101.96553 	&	0.674 	&$	237 	^{+	5.1 	}_{ -	3.7 	}$&$	231 	^{+	34.7 	}_{ -	17.0 	}$	\\
 &	2458101.97445 	&	0.693 	&$	243 	^{+	4.1 	}_{ -	3.2 	}$&$	237 	^{+	33.7 	}_{ -	18.0 	}$	\\
  \hline
CFHT-ESPaDOnS & 2459808.11664 & 0.750 & $ 272^{+3.0}_{-3.0}$ & --- & \\
 & 2459808.13110 & 0.780 & $ 267^{+3.0}_{-3.0}$ & --- &\\
\hline
\end{tabular}
\end{center}
\end{table*}

We determine the radial velocities of absorption lines and the prominent 
H$\alpha$ emission line (in the following wavelength window [$6520$ \AA, 
$6620$ \AA]), respectively. The radial-velocity measurements are listed in 
Table~\ref{tbl:rv}. 

\subsubsection{Inferring the Rotational Broadening Velocity $V_{\mathrm{rot}}\sin i$}
\label{sect:vsini}
We can measure the rotational broadening velocity $V_{\mathrm{rot}}\sin i$ 
of the visible star from the CFHT spectra of the visible star. Note that the 
spectral resolution of the \textit{PyHammer} template 
is too low to measure $V_{\mathrm{rot}}\sin i$. We determine the 
best-matching rotating template from the Marcs \citep{Gustafsson2008} 
theoretical atmospheric model. Our procedures are as follows. 
First, the radial velocities of the two CFHT spectra 
are measured with the aforementioned method. Note that the radial velocities 
are consistent with the LAMOST measurements at the same orbital phases. 
Second, the two CFHT spectra are shifted to have zero radial velocity with 
respect to the Marcs best-matching models and co-added to increase the 
signal-to-noise ratio. Third, we generate the Marcs spectra with 
$T_{\rm eff} = 4100$ K and $\log{g} = 4.66$ (see Section~\ref{sect:sed-data}), 
but different $[\mathrm{Fe/H}]$, from $-1.0$ to 
$0.1$ in a $0.1$ step size. These Marcs spectra are downgraded to the 
spectral resolution of $R=68, 000$. Fourth, the best-fit template is 
determined by minimizing $\chi^{2}=\sum (f_{\mathrm{o, \lambda}} - 
f_{\mathrm{m, \lambda}})^2/\sigma^2_{\mathrm{o, \lambda}}$, 
where $f_{\mathrm{o, \lambda}}$, $\sigma_{\mathrm{o, \lambda}}$, and 
$f_{\mathrm{m, \lambda}}$ are the co-added CFHT flux, flux uncertainty, 
and the Marcs model flux, respectively. The best-matching template 
has $[\mathrm{Fe/H}]=-0.3$. 

\begin{figure*}
  \centering
  \includegraphics[keepaspectratio, width=0.8\textwidth]{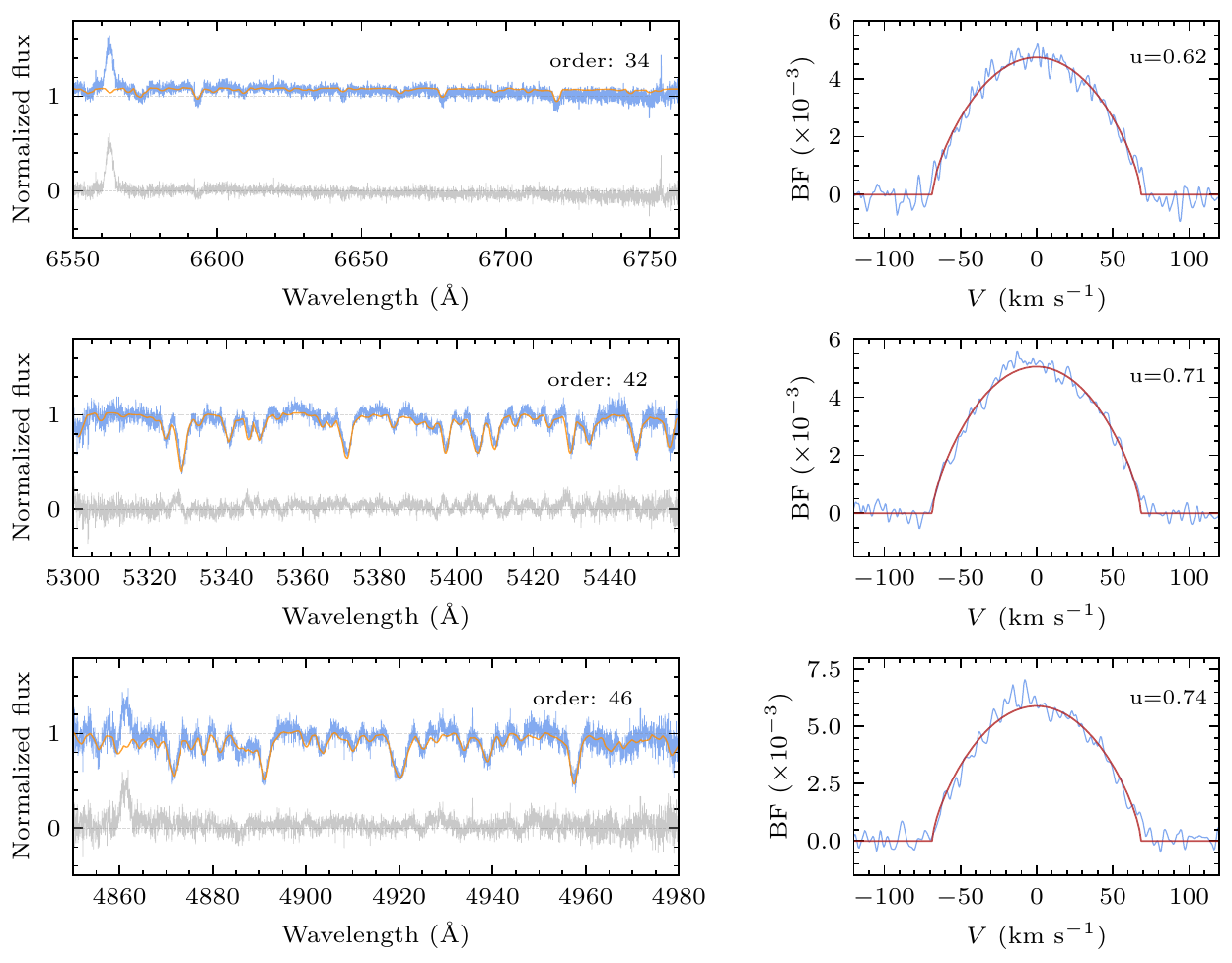} \\
  \caption{\textbf{The CFHT spectra and rotational broadening measurements.} Left panels: 
  the co-added CFHT spectra (blue) at echelle orders 34 (top), 42 (middle), and 46 
  (bottom). These orders are selected from the three wavelength windows (see 
  text), respectively. The best-fitting rotational-broadened Marcs template 
  is shown in orange, and the residuals are shown in grey. The H$\alpha$ 
  (6564 \AA) and H$\beta$ (4863 \AA) emission lines are clearly seen in orders 
  34 and 46, which are masked when measuring the $V_{\rm rot} \sin{i}$. Right 
  panels: the three best-fitting broadening functions (blue) and rotational 
  profiles (red) for the three wavelength windows. }
  \label{fig:vsini}
\end{figure*}

\begin{figure*}
  \centering
  \includegraphics[keepaspectratio, width=0.8\textwidth]{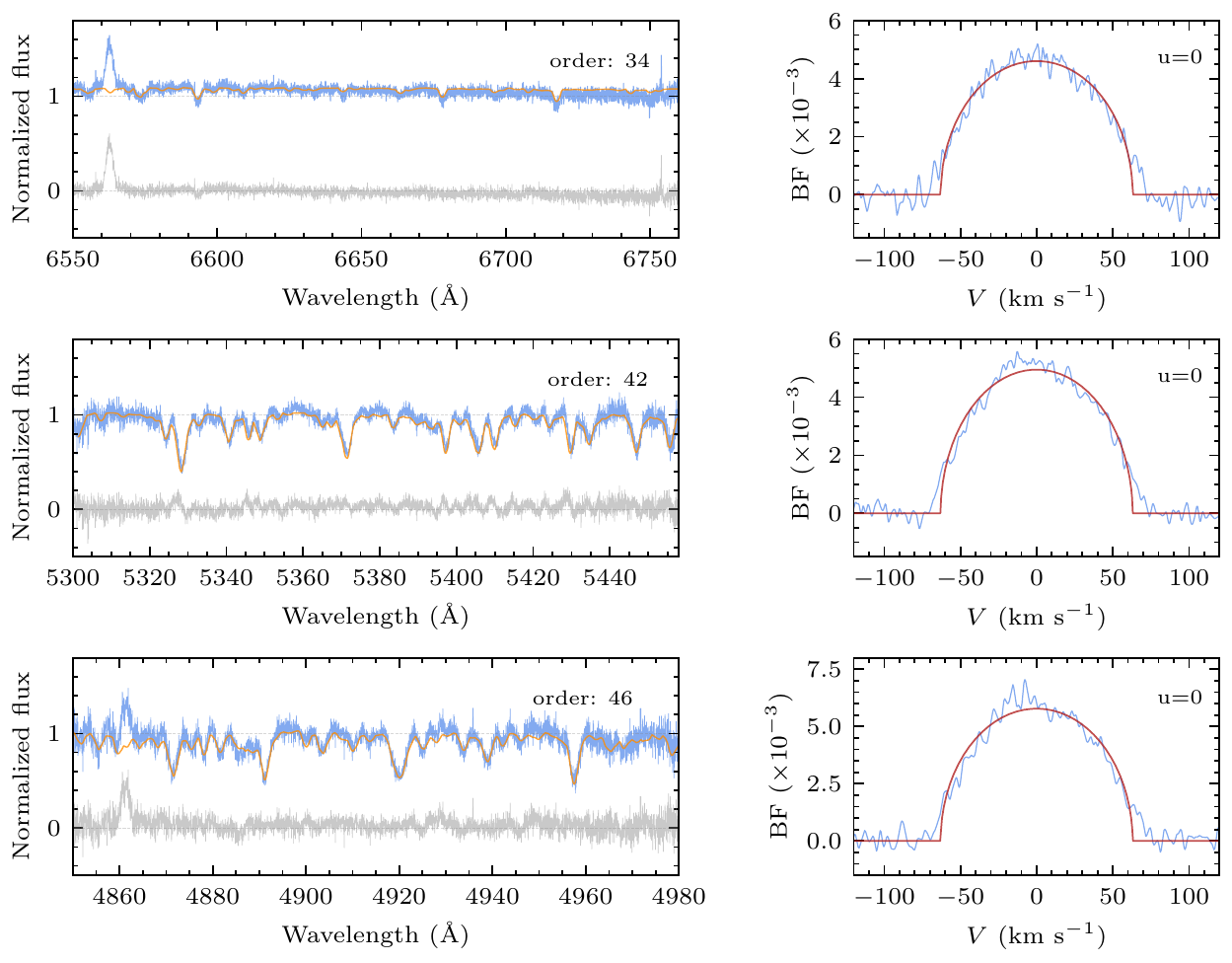} \\
  \caption{The same as Figure~\ref{fig:vsini}, but without limb darkening for 
  absorption lines. }
  \label{fig:vsini_sd}
\end{figure*}

We used the broadening function \citep{Rucinski1992} to measure the 
$V_{\mathrm{rot}}\sin i$. The steps are as follows.  
\begin{itemize}
  \item Wavelength window selection. We choose three wavelength windows, 
  each with 4--6 echelle orders, to measure 
  $V_{\mathrm{rot}}\sin i$. The orders are: orders 
  50--45, orders 44--39, and orders 37--34, which cover three wavelength 
  windows: 446.1--509.8 nm, 505.9--589.4 nm, and 559.8--667.5 nm, respectively.
  These orders contain many photospheric absorption lines and are free of 
  telluric absorption lines. In addition, we mask Balmer emission lines. We 
  use the average of the three best-fitting $V_{\mathrm{rot}}\sin i$ values 
  from the three wavelength windows as our final $V_{\mathrm{rot}}\sin i$. 

  \item Broadening function. We construct a design matrix $\hat{D}$ using the 
  best-fit, non-broadened Marcs template, denoted as $T$ \citep{Rucinski1992}. 
  Each column of the 
  $\hat{D}$ contains a continuum-normalized $T$ with a specific radial velocity $v$, 
  spanning from $-$150 $\mathrm{km~s^{-1}}$ to 150 $\mathrm{km~s^{-1}}$. We aim to 
  solve the equation $\hat{D} \vec{B} = \vec{O}$, where $\vec{B}$ and $\vec{O}$ are 
  the unknown broadening function (BF) and the continuum-normalized, co-added CFHT 
  spectrum. The singular value decomposition method is implemented \citep{Rucinski1992} 
  to solve the equation and constrain the BF. 

  \item Rotational broadening. The spectral broadening contains the rotational 
  broadening, the instrumental broadening ($\sim$4.4 $\mathrm{km~s^{-1}}$), and 
  the broadening by 
  macro-turbulence ($\simeq 2\ \mathrm{km\ s^{-1}}$). The latter two are small and 
  negligible with respect to the rotational broadening. The 
  $V_{\mathrm{rot}}\sin{i}$ is obtained by minimizing the $\chi^2$ between the 
  calculated BF and the rotational profile \citep[Equ. 18.14]{Gray2005}: 
  \begin{equation}
  \begin{aligned}
  G(x) =  
  & \frac{2(1-\epsilon)(1-x^2)^{1/2}+\frac{\pi\epsilon}{2} (1-x^2)}{\pi 
  (1-\frac{\epsilon}{3})} & ~\text{for}~\lvert x \rvert <1 \\ 
  & 0 & ~\text{for}~\lvert x \rvert>1 \\
  \end{aligned} \, ,
  \end{equation}
  where $x= V / (V_{\mathrm{rot}}\sin{i})$ and $V$ and $\epsilon$ are the shift velocity 
  due to rotation and the limb-darkening coefficient, respectively. For the limb-darkening 
  coefficient, we consider two cases: first, the limb-darkening 
  coefficient for the absorption lines is zero; second, the limb-darkening coefficient 
  is the same as for the continuum \citep[as given by][]{Claret2011, Claret2017}. The real case is 
  probably somewhere between the two extremes \citep{Shahbaz2003}. 
  The best-fitting result is shown in Figures~\ref{fig:vsini} and~\ref{fig:vsini_sd}. 

  \item Uncertainty estimation. We use the Monte Carlo simulation to estimate the 
  uncertainty of $V_{\mathrm{rot}}\sin{i}$. The best-fitting broadened Marcs template 
  is randomly perturbed according to the observational flux uncertainties. We measure 
  the mock $V_{\mathrm{rot}}\sin{i}$ of the perturbed template. This process is 
  repeated $5,000$ times. The 16th-percentile and the 84-percentile of 
  the mock $V_{\mathrm{rot}}\sin{i}$ distribution are used as the lower and the 
  upper error bars. 
\end{itemize}

For the first case (i.e., the limb-darkening coefficient of the absorption lines is zero), 
$V_{\mathrm{rot}}\sin{i}=63^{+8}_{-6}\ \mathrm{km\ s^{-1}}$; for the second case (i.e., 
absorption lines and continua share the same limb-darkening coefficient) 
$V_{\mathrm{rot}}\sin{i}=68.37^{+0.06}_{-0.09}\ \mathrm{km\ s^{-1}}$. 

\subsection{Optical-to-infrared light curves}
\label{sect:lc}
J2354 has high-cadence (i.e., every 30 minutes) 30-day-long TESS full-frame 
images in Sector 17 \citep{Ricker2015, Stassun2019}. The TESS 
Sector 17 light curve is extracted and stored in STScI MAST 
(\url{https://mast.stsci.edu/portal/Mashup/Clients/Mast/Portal.html}). 
The STScI MAST PDC-SAPFLUX and its associate measurement errors (in units 
of $\mathrm{e^{-}\ s^{-1}}$) are used to construct the TESS light curve. 

J2354 is also monitored by several ground-based time-domain surveys, 
namely, the All-Sky Automated Survey for Supernovae \citep[ASAS-SN;][]{ASASSN} 
and the Zwicky Transient Facility \citep[ZTF;][]{ZTF}. The ASAS-SN 
light curve is retrieved from ASAS-SN Variable Stars Database 
(\url{https://asas-sn.osu.edu/variables}) and the corresponding cadence is 
$2\sim 3$ days; the ZTF $g$ and $r$ data are downloaded via 
IRSA 
(\url{https://irsa.ipac.caltech.edu/cgi-bin/Gator/nph-scan?mission=irsa&submit=Select&projshort=ZTF}) 
whose cadences are about $1$ day.  

The Wide-field Infrared Survey Explorer \citep[WISE;][]{WISE,NEOWISE} 
can provide WISE $W1$ and $W2$ infrared light curves for J2354; their 
cadences are $\sim 1.5$ days. We only consider observations with 
\texttt{ccflags} is 0; in this case, the target is unaffected by known artifacts.

\subsection{X-ray and UV observations}
\label{sect:xray-uv}
We requested six Swift (the Neil Gehrels Swift Observatory) ToO observations 
(see Table~\ref{tbl:swift}) to investigate the possible X-ray and UV variability 
in J2354 (Target ID: $15376$) utilizing the X-ray Telescope (XRT) and 
Ultraviolet/Optical Telescope (UVOT). 

We extract the Swift UVOT light curves via the \texttt{uvmaghist} tool in 
\textsc{HEASoft} 
version 6.30 (\url{https://heasarc.gsfc.nasa.gov/docs/software/heasoft/}). 
The source region is selected using a circle with a radius of $5$ arcsec centred 
on the optical position, and the corresponding background region is selected by 
an annulus with a $12.5$ arcsec inner radius and a $25.0$ arcsec outer radius. 

X-ray observations, which were taken in the photon counting (PC) mode, show a 
low and highly variable count rate. We process the data with the packages 
available in \textsc{HEASoft}, and start with the standard data screening by 
using the task \texttt{xrtpipeline}. The source events are extracted from a 
circle of radius 15 pixels centred at the source position, having 
26, 5, 13, 3, 1, and 3 photons (without background subtraction) for each 
individual observation. Hence, the X-ray emission of J2354 is highly variable 
and can only be detected in the first and third observations. 

For two detections, we further extract the background spectra from a circular 
region of radius 30 pixels away from source. The ancillary response files are 
generated with the task \texttt{xrtmkarf}, and the response file (v014) from 
the CALDB database is adopted for the subsequent spectral analyses. The 
spectra are grouped to have at least 2 counts per bin, and they are analysed 
with \textsc{xspec} version 12.12.1 \citep{xspec}. Because of limited photons, 
we fix the neutral hydrogen column density to the Galactic absorption 
($5.65 \times 10^{20}$ cm$^{-2}$) towards the direction of the source 
\citep{HI4PI2016}, and change the fit statistic to C-statistic. Fitting the 
spectra in 0.3--2.5 keV with an absorbed power-law model\footnote{We 
also used other models, e.g., the blackbody model or the summation of 
the blackbody and powerlaw model. Due to the limited number of photon 
numbers, these models are statistically indistinguishable. } 
(\texttt{tbabs*powerlaw} in \textsc{xspec}), we obtain the best fitted 
parameters and their uncertainties in 90\% confidence level as follows: 
for the first observation, the photon index $\Gamma = 2.9_{-0.7}^{+0.8}$, 
the 0.3--2 keV unabsorbed flux $f = 1.2_{-0.4}^{+0.6} \times 10^{-12}$ 
erg~cm$^{-2}$~s$^{-1}$, and C-Statistic/dof$= 9.0/10$; for the third 
observation, $\Gamma = 2.7_{-2.6}^{+2.7}$, 
$f = 3.4_{-2.2}^{+19.4} \times 10^{-13}$ erg~cm$^{-2}$~s$^{-1}$, and 
C-Statistic/dof$= 9.4/3$. 

\begin{table*}[!t]
  \begin{center}
    \caption{The Swift observational log. \label{tbl:swift}}
      \begin{tabular}{cccccr}
      \hline
      Epoch  & Observation date & XRT mode & X-ray detection & UVOT mode & AFST 
      (second) \\
      \hline 
      (1) & (2) & (3) & (4) & (5) & (6) \\ 
      \hline
      1	& 2022-10-24 & PC & Yes & 0x30ed & 1150 \\
      2	&	2022-11-02 & PC & No & 0x308f & 1490 \\
      3	&	2022-11-06 & PC & Yes & 0x308f & 1570 \\
      4	&	2022-11-12 & PC & No & 0x308f & 440 \\
      5 & 2022-11-16 & PC & No & 0x308f & 595 \\
      6 & 2022-11-19 & PC & No & 0x308f & 755 \\
      \hline
      \end{tabular}
      \\
      \textbf{Notes.} AFST refers to the AS-Flown Science Timeline. Each of the 
      first, second, and third observation consists 
      of two or three separate exposures. The UVOT fluxes are independently 
      measured in each exposure.  
  \end{center}
\end{table*}

\subsection{The UV-to-infrared spectral energy distribution}
\label{sect:sed-data}
The multi-band (i.e., from UV to infrared) spectral energy distribution (SED) 
of J2354 is collected from GALEX \citep{Martin2005, Morrissey2007}, 
the Sloan Digital Sky Survey \citep[SDSS;][]{Abazajian2009}, the AAVSO Photometric 
All-Sky Survey \citep[APASS;][]{Henden2016}, the Panoramic Survey Telescope and 
Rapid Response System \citep[Pan-STARRS;][]{Chambers2016, Magnier2020a, 
Magnier2020b, Magnier2020c, Waters2020}, Gaia \citep{Gaia}, TESS \citep{Stassun2019}, 
the Two Micron All-Sky Survey \citep[2MASS;][]{Skrutskie2006, Cutri2003a} 
and the Wide-field Infrared Survey Explorer \citep[WISE;][]{WISE} (see 
Table~\ref{tbl:sed}). 

\begin{figure*}
  \centering
  \includegraphics[keepaspectratio, width=0.5\textwidth]{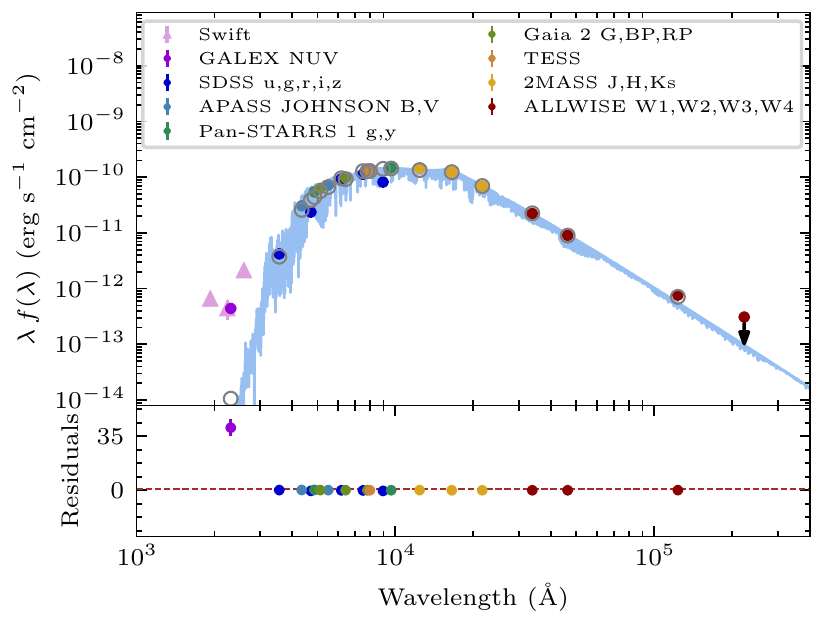} \\
  \caption{\textbf{The SED fitting result.} 
  Upper: the best-fitting template (the blue curve) versus the 
  observed SED (the filled dots). The grey open circles represent the model 
  fluxes. The best-fitting 
  stellar template (i.e., the blue curve) matches well with the observed optical 
  SED. For the NUV band ($\lambda=2305$ \AA) and the three 
  \texttt{Swift} UV bands (UVW2, UVM2, UVW1), the stellar template 
  fluxes are significantly dimmer than the observed ones. 
  This NUV excess is due to the chromospheric activities, which are not taken 
  into account when creating the stellar template. 
  Lower: the ratios of the differences between the observed and model fluxes 
  to the model fluxes. Note that the NUV, W3, and W4 fluxes are not considered 
  during the SED fitting procedure. In addition, $4\%$ systematic uncertainties 
  are added when fitting the SED.}
  \label{fig:sed}
\end{figure*}

We use the single-star model, \textit{ARIADNE} (spectrAl eneRgy dIstribution bAyesian 
moDel averagiNg fittEr; \url{https://github.com/jvines/astroARIADNE}; \cite{ARIADNE}) 
to fit the observed optical-to-infrared SED and determine the model-free parameters: 
$T_{\mathrm{eff}}$, log $g$, [Fe/H], $D$, and $R$. Because of the close distance, 
a negligible extinction value of $A_{\rm{v}} = 0.00^{+0.03}_{ -0.00}$ mag is derived 
by using the colour excess $E(g-r) = 0.00^{+0.01}_{-0.00}$ from 3D Dust Mapping 
(\url{http://argonaut.skymaps.info/}). As a result, the extinction value is 
fixed to be zero during the SED fitting. The prior distribution 
for the distance ($D$) is a normal distribution whose mean and standard deviation 
are fixed according to the \textit{Gaia} result. The prior distribution for 
[Fe/H] is a normal distribution whose mean and standard deviation 
are $-0.3$ (Section~\ref{sect:vsini}) and 0.3, respectively. \textit{ARIADNE} uses 
the stellar evolution model (isochrones) to resolve the parameter degeneracy. 
The adopted stellar synthetic SED models are Phoenix v2 
(\href{ftp://phoenix.astro.physik.uni-goettingen.de/HiResFITS/PHOENIX-ACES-AGSS-COND-2011/}{Phoenix v2 link}), 
BT-Models (\href{http://osubdd.ens-lyon.fr/phoenix/}{BT-Models link}), 
Castelli \& Kurucz (\url{http://ssb.stsci.edu/cdbs/tarfiles/synphot3.tar.gz}), 
and Kurucz 1993 (\url{http://ssb.stsci.edu/cdbs/tarfiles/synphot4.tar.gz}). 
\textit{ARIADNE} uses nested sampling algorithms to estimate the best-fitting 
free parameters. Consistent with its single-lined nature, the best-fitting 
single-star SED model fits the observed optical-to-infrared data well 
(Figure~\ref{fig:sed}). 

\begin{table*}[h!]
  \centering
  \caption{The multi-band SED. \label{tbl:sed}}
  \begin{tabular}{ccrrcrc}
  \hline
  Telescope  & Band & $\lambda_{\rm central}$  & magnitude & system & 
  $\lambda f (\lambda)$ & Ref. \\ 
  \hline
             &   &  \AA &  mag &  & $\mathrm{\times 10^{-13}erg\ s^{-1}\ cm^{-2}}$  & \\ 
  \hline
 GALEX &	NUV	&	2305 	&	20.1 	$\pm$	0.14 	&	AB	&	4.4 	$\pm$	0.59	
	&\citep{Martin2005,Morrissey2007}\\
  \hline
 SDSS &	u	&	3562 	&	17.16 	$\pm$	0.01 	&	AB	&	42 	$\pm$	0.38
	& \citep{Abazajian2009}	\\
&	g	&	4719 	&	14.974 	$\pm$	0.006 	&	AB	&	236 	$\pm$	1.3 & \citep{Abazajian2009}	\\
&	r	&	6186 	&	13.185 	$\pm$	0.003 	&	AB	&	934 	$\pm$	2.6 & \citep{Abazajian2009}	\\
&	i	&	7500 	&	12.767 	$\pm$	0.001 	&	AB	&	1132 	$\pm$	1.0 & \citep{Abazajian2009}	\\
&	z	&	8961 	&	12.932 	$\pm$	0.008 	&	AB	&	816 	$\pm$	6.1 & \citep{Abazajian2009}	\\
  \hline
 APASS & B	&	4348 	&	14.9 	$\pm$	0.13 	&	Vega	&	306.50 	$\pm$	37	& 
\citep{Henden2016} \\
 & V	&	5505 	&	13.61 	$\pm$	0.017 	&	Vega	&	721 	$\pm$	11 & \citep{Henden2016}	\\
  \hline
 Pan-STARRS &	g	&	4866 	&	14.036 	$\pm$	0.005 	&	AB	&	545 	$\pm$	2.6	& 
\citep{Chambers2016,Magnier2020a,Waters2020,Magnier2020b, Magnier2020c} \\
&	y	&	9633 	&	12.215 	$\pm$	0.001 	&	AB	&	1474 	$\pm$	1.4 & 
\citep{Chambers2016,Magnier2020a,Waters2020,Magnier2020b,Magnier2020c}	\\
  \hline
 Gaia &	BP	&	5129 	&	13.792 	$\pm$	0.009 	&	Vega	&	631 	$\pm$	5.2 & \citep{Gaia}\\
&	G	&	6425 	&	13.037 	$\pm$	0.002 	&	Vega	&	977 	$\pm$	1.8 & \citep{Gaia}	\\
&	RP	&	7800 	&	12.194 	$\pm$	0.006 	&	Vega	&	1334 	$\pm$	7.4 & \citep{Gaia}	\\
  \hline
 TESS &	red	&	7972 	&	12.261 	$\pm$	0.008 	&	Vega	&	1331 	$\pm$	9.8	&
  \citep{Stassun2019}\\
  \hline
 2MASS &	J	&	12408 	&	11.06 	$\pm$	0.02 	&	Vega	&	1452 	$\pm$	27
	& \citep{Skrutskie2006,Cutri2003a}	\\
&	H	&	16514 	&	10.45 	$\pm$	0.019 	&	Vega	&	1235 	$\pm$	21 & \citep{Skrutskie2006,Cutri2003a}	\\
&	Ks	&	21656 	&	10.32 	$\pm$	0.016 	&	Vega	&	691 	$\pm$	10 & \citep{Skrutskie2006,Cutri2003a}	\\
  \hline
 ALLWISE &	W1	&	33792 	&	10.24 	$\pm$	0.023 	&	Vega	&	223 	$\pm$	4.7	&  \citep{Cutri2021} \\
&	W2	&	46293 	&	10.23 	$\pm$	0.02 	&	Vega	&	91 	$\pm$	1.7 &  \citep{Cutri2021}	\\
&	W3	&	123340 	&	10.06 	$\pm$	0.056 	&	Vega	&	7.7 	$\pm$	0.4 &  \citep{Cutri2021}	\\
&	W4	&	222530 	&	$>$ 8.89			&	Vega	&	$<$3.09		&  \citep{Cutri2021}	\\
  \hline
\end{tabular}
\\
\textbf{Notes.} Photometric measurements with zero uncertainties or photometric flags (e.g., 
several Pan-STARRS bands) are excluded. Additional $4\%$ systematic uncertainties (i.e., 
the semi-amplitude of TESS flux variations) are added when performing the SED fitting. 
\end{table*}

\begin{figure*}
  \centering
  \includegraphics[keepaspectratio, width=0.8\textwidth]{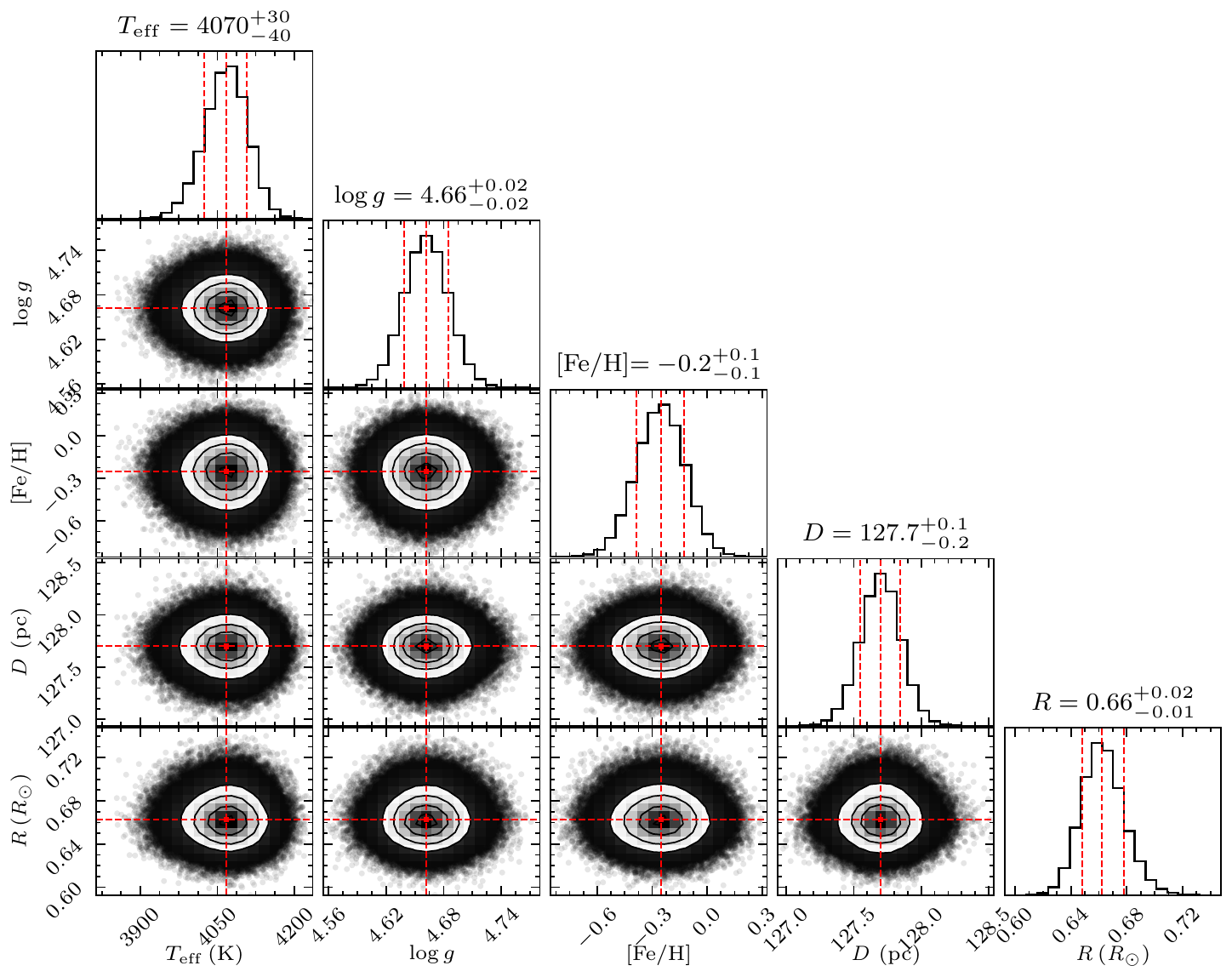} \\
  \caption{\textbf{The posterior distributions of the stellar parameters derived from the 
  ARIADNE SED fitting code.} The dashed vertical lines indicate the 16th, 50th, 
  and 84th percentiles of the distributions. The contours indicate the joint 
  distributions of two parameters. Note that the ARIADNE SED fitting results 
  are consistent with the values obtained from the LAMOST and Gaia 
  databases. }
  \label{fig:pms}
  \end{figure*}

\textit{ARIADNE} uses nested sampling algorithms to obtain the posterior distributions 
of the free parameters. Hence, we obtain $T_{\mathrm{eff}} = 4070 ^{ +30}_{ -40}$ K, 
$\log g = 4.66 \pm 0.02$ dex, radius $R = 0.66^{+0.02}_{-0.01}\ R_{\odot}$ (see 
Figure~\ref{fig:pms}), and the 
bolometric luminosity $L_{\mathrm{bol}} = 4\pi R^{2}\sigma 
T_{\mathrm{eff}}^{4} = 0.108\pm 0.005\ L_{\odot}$. 

The stellar parameters are derived in different ways to check for consistency. 
First, the effective temperature $T_{\rm{eff}} = 4171 \pm 36$ K and 
$\log g = 4.5 \pm 0.12$ are reported in the LAMOST LDR8 database. Moreover, 
the $V$-band flux (13.61 mag) from UCAC4 \citep{Zacharias2013}, the \textit{Gaia} 
distance (127.7 $\pm$ 0.3 pc), the extinction value 
$A_{\rm v} = 0.00 ^{ +0.03}_{-0.00}$ mag from 3D Dust Mapping, and the 
empirical bolometric correction \citep{Torres2010} can be used to derive the 
bolometric luminosity $L_{\rm{bol}} = 0.105\pm 0.008\,L_{\odot}$. The 
corresponding radius 
$R_{\mathrm{vis}} = \sqrt{L_{\rm{bol}}/(4\pi \sigma T_{\mathrm{eff}}^{4})} 
= 0.62\pm 0.03\,R_{\odot}$. Second, as mentioned above, the tool 
\textit{ARIADNE} can be 
used to fit the observed SED and infer the physical parameters of our source 
(see Figure~\ref{fig:pms}). Third, we use the empirical relations of colour 
indexes and radius for main-sequence stars \citep{Boyajian2012} to derive 
$R_{\mathrm{vis}}$, which is from $0.62\sim 0.65\ R_{\odot}$. The parameters 
from \textit{ARIADNE} are broadly consistent with those from other methods. 

\section{Result}
\label{sect:result}
\subsection{The Orbital Parameters and Ephemeris}
\label{sect:mlcs}

The high-cadence (30 minutes) but short-duration ($\sim 30$ days) TESS light 
curve shows evident periodic signatures. We calculate the corresponding 
power spectral density (PSD) via the Lomb-Scargle algorithm \citep{Scargle1981}. 
Indeed, the Lomb-Scargle periodogram has two unequivocal peaks; the first and 
second peaks correspond to $P_1=0.23996$ days and $P_2=0.47991$ days, 
respectively. One of the two periodic signatures may be related to the 
orbital modulation. 

\begin{figure*}
  \centering
  \includegraphics[keepaspectratio, width=0.5\textwidth]{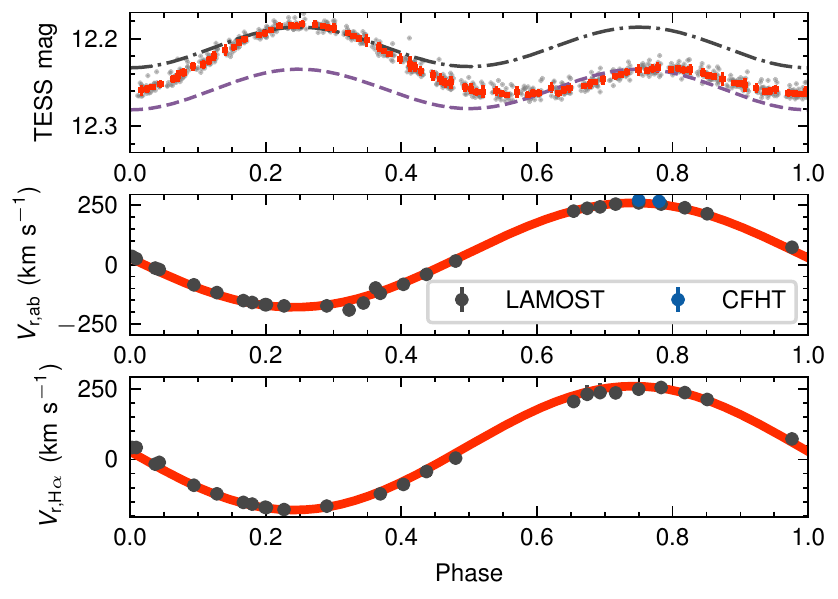}
  \caption{\textbf{The phase-folded TESS light curve and radial velocities of J2354.} 
  Upper panel: the phase-folded (with $P_{\mathrm{orb}}=0.47992$ days) TESS 
  light curves for J2354. The grey points and red squares correspond to the 
  TESS observations in Sector 17 and its running mean over ten data points, 
  respectively. The purple dashed and black dot-dashed curves represent the 
  ellipsoidal-modulation light curves normalized according to the TESS magnitude 
  at $\Phi=0.75$ and $\Phi=0.25$, respectively. The model parameters are 
  $i=73$ degrees, $M_{\mathrm{vis}}=0.73\ M_{\odot}$, and 
  $M_{\mathrm{inv}}=1.4\ M_{\odot}$. The TESS light curve shows 
  additional variations beyond ellipsoidal modulations. 
  Middle panel: the phase-folded LAMOST (black dots) and CFHT (blue dots) 
  radial-velocity measurements of stellar absorption lines. The radial-velocity 
  variations can be perfectly fitted by a sinusoidal function (the red curve), 
  with the semi-amplitude $K_{\mathrm{vis}}=219.4\pm 0.5\ 
  \mathrm{km\ s^{-1}}$. 
  Lower panel: the phase-folded LAMOST H$\alpha$ radial velocities (black dots), 
  which vary in tandem with the radial velocities of stellar absorption 
  lines (the red curve). Hence, H$\alpha$ is produced by the chromospheric 
  activity of the visible star. Note that the error bars of the data are 
  small and invisible.} \label{fig:lc}
\end{figure*}

The orbital period can be further determined by fitting the radial-velocity 
variations. A Python code \textit{The Joker} 
(\url{https://github.com/adrn/thejoker}; \citep{Price-Whelan2017}), 
is used to fit a two-body system to the radial velocities (Figure~\ref{fig:lc}). 
In the two-body system, the radial velocities are 
$V_{\mathrm{r}}(t) = \gamma + K (\cos(f + \omega) + 
e\cos(\omega))$, where $\gamma$, $K$, $f$, $\omega$, and $e$ 
represent the long-term mean barycentre velocity, the semi-amplitude, 
the true anomaly, the argument of periastron, and eccentricity, respectively. 
Note that the true anomaly $f$ is a function of $P_{\mathrm{orb}}$. An 
additional overall noise term ($s$) is allowed to account for the possible 
underestimation of the uncertainties of $V_{\mathrm{r}}$ \citep{Price-Whelan2017}. 
We also assume that the LRS and MRS $V_{\mathrm{r}}$ 
measurements might have a small constant 
velocity offset ($dv_0$). The prior distribution for the standard deviation of 
the additional noise term $s$ is assumed to be a log-normal one (whose mean and 
standard deviation are $0.5\ \mathrm{km\ s^{-1}}$ and $1.5\ \mathrm{km\ s^{-1}}$, 
respectively). The prior distribution for $dv_0$ is a normal one (whose mean and 
standard deviation are $0\ \mathrm{km\ s^{-1}}$ and $5\ \mathrm{km\ s^{-1}}$). 
The prior distribution for $P_{\mathrm{orb}}$ is a uniform one in 
[$0.1$ days, $0.7$ days]. The prior distributions for $e$, 
$\omega$, $K$, and $\gamma$ are set to the default ones, i.e., 
a beta distribution for $e$, a uniform distribution for $\omega$, a normal 
distribution for $K$, and a normal distribution for $\gamma$, respectively. 
\textit{The Joker} uses the MCMC algorithm to sample the posterior 
distributions of model parameters. The best-fitting 
$P_{\mathrm{orb}}$ is $0.47992$ days with the $1\sigma$ statistical uncertainty 
(the reported uncertainties always refer to the $1\sigma$ ranges, unless 
otherwise specified) of $10^{-5}$ days. The other period ($0.23996$ days) 
is incompatible with the radial-velocity variations. Hence, we conclude that 
the orbital period of J2354 is $P_{\mathrm{orb}}=0.47992$ days. 

The ephemeris of the system is 
\begin{equation}
  {\rm T}(\phi = 0) =2459130.110499\,{\rm HJD} 
  + 0.47992\times N\ \,
\end{equation}
where the phase $\phi = 0$ corresponds to the 
visible star in the superior conjunction, and HJD is the Heliocentric Julian 
Date. Multi-band light curves from various telescopes can be phase-folded 
according to the same orbital ephemeris. The phase-folded light curves are 
asymmetric (Figure~\ref{fig:lc}), driven by both ellipsoidal modulations 
(due to the tidal distortion 
of the visible star) and stellar activities (presumably due to e.g., 
starspots; for a detailed discussion, see Section~\ref{sect:variation}). 
The ellipsoidal modulations account for the 
$P_1=0.23996$-day TESS PSD peak. 

Other orbital parameters are determined by the radial-velocity fitting. 
For instance, The eccentricity $e = 0.002\pm 0.002$, i.e., the 
orbit is nearly circular; the systematic velocity of J2354 is $\gamma = 
41^{+2.4}_{-2.3}\ \mathrm{km\ s^{-1}}$. 
The radial-velocity semi-amplitude of the absorption lines and the H$\alpha$ 
emission line are $K_{\mathrm{vis}}=219.4\pm 0.5\ \mathrm{km\ s^{-1}}$ and 
$K_{\mathrm{H\alpha}}=216\pm 2\ \mathrm{km\ s^{-1}}$. Hence, the mass 
function is $f(M_{\mathrm{inv}})=0.525\pm 0.004\ M_{\odot}$.

\subsection{The Mass Function of the Compact Object}
\label{sect:fmass}

The mass function of the invisible object in the binary system is
\begin{equation}
  f(M_{\rm{inv}}) = 
(M_{\rm{inv}}\sin i)^{3}/(M_{\rm{vis}}+M_{\rm{inv}})^{2} = 
K_{\rm{vis}}^{3}P_{\rm{orb}}/(2\pi G)\ \,
\end{equation}
where $M_{\rm{inv}}$, $M_{\rm{vis}}$, and $i$ are the mass of the 
invisible object, the mass of the visible star, and the inclination 
angle, respectively. According to the radial-velocity fitting results, 
the mass function of the invisible object in J2354 is 
$f(M_{\rm{inv}})=0.525\pm 0.004\ M_{\odot}$, which is also the 
lower limit for $M_{\rm{inv}}$. 

Strong constraints on $M_{\rm{inv}}$ can be obtained if one knows 
$M_{\rm{vis}}$. We adopt $\log g$ and $R_{\mathrm{vis}}$ derived from 
\textit{ARIADNE} to calculate $M_{\rm{vis}}$ and its uncertainty because 
$M_{\mathrm{vis}}^g=gR_{\mathrm{vis}}^2/G$ (where $G$ is the gravitational 
constant). Note that the \textit{ARIADNE} code utilizes 
the stellar evolution model (isochrones) to significantly improve the 
constraint of $\log g$. Ten thousand realizations 
of $\log g$ and $R_{\mathrm{vis}}$ from the \textit{ARIADNE} sampling results 
are used to construct the distribution of $M_{\rm{vis}}^g$. The 
median $M_{\mathrm{vis}}^g$ and its $1\sigma$ uncertainty (estimated via the 
16th and 84th percentiles) are $0.73^{+0.06}_{-0.05}\ M_{\odot}$. Alternatively, the 
mass-luminosity relation \citep[MLR;][]{Henry1993} and the observed absolute 
magnitudes in J, H, and K bands (see Table~\ref{tbl:sed}) can be used to 
estimate $M_{\mathrm{vis}}$; the resulting $M_{\mathrm{vis}}$ from the three 
bands are similar, i.e., $0.66_{-0.09}^{+0.11}\ M_{\odot}$. The MLR-based 
$M_{\mathrm{vis}}$ are statistically consistent (within $1\sigma$ uncertainties) 
with the SED-fitting $\log g$-based stellar mass. The SED-fitting $\log g$-based 
stellar mass and radius of J2354 is also statistically consistent with the 
empirical mass-radius relation for main sequence stars \citep{Boyajian2012}. 

Given $f(M_{\rm{inv}})$ and $M_{\rm{vis}}$, $M_{\rm{inv}}$ anticorrelates 
with the inclination angle $i$. For the perfectly edge-on case (i.e., 
$i=90$ degrees), the corresponding mass of the invisible object is 
$1.29 \pm 0.04\ M_{\odot}$. Hence, $M_{\rm{inv}}\geq 1.29 \pm 0.04\ 
M_{\odot}$. If the unseen companion is a normal star, it would outshine 
the visible component. Hence, the unseen companion in J2354 ought to be 
a near-Earth compact object. 

\begin{figure*}
  \centerline{\includegraphics[keepaspectratio, width=0.5\textwidth]{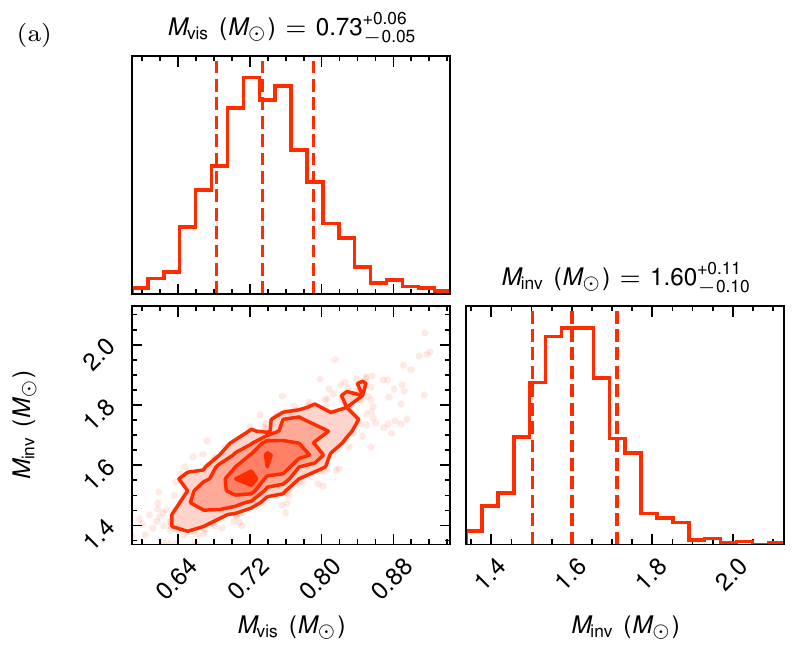}
  \includegraphics[keepaspectratio, width=0.5\textwidth]{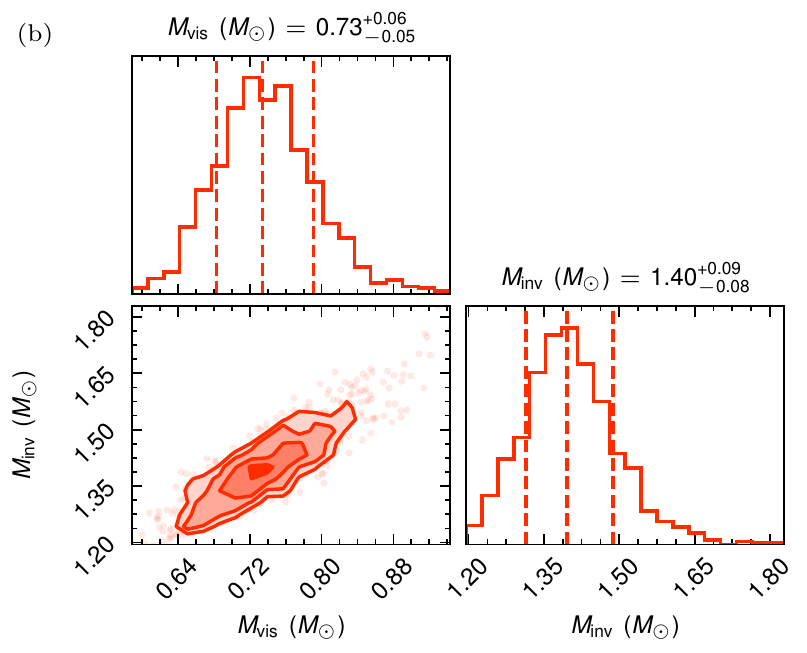}}
    \caption{\textbf{The joint probability density distribution (PDF) of the mass of the 
    invisible object ($M_{\mathrm{inv}}$) versus that of the visible star 
    ($M_{\mathrm{vis}}$).} The left and right panels are for the limb darkening 
    coefficient of absorption lines fixed to be zero and the continuum value, 
    respectively. For each panel, the two histograms indicate the PDFs of 
    $M_{\mathrm{inv}}$ 
    and $M_{\mathrm{vis}}$; in each histogram, the three dashed vertical lines 
    illustrate the $16$-th, $50$-th, and $84$-th percentiles of the distribution. }
    \label{fig:mass}
\end{figure*}

\subsection{The Mass of the Compact Object}
\label{sect:minv}

To determine $M_{\rm{inv}}$, we should estimate the inclination angle $i$. 
The rotational broadening velocity of the visible star, which is measured 
from the two high-resolution CFHT/ESPaDOnS spectra, depends upon the limb 
darkening coefficient of absorption lines: if the coefficient is zero 
(case I), $V_{\mathrm{rot}}\sin i =63^{+8}_{-6}\ \mathrm{km\ s^{-1}}$; if 
the coefficient is the same as for the continua (case II), 
$V_{\mathrm{rot}}\sin i =68.37^{+0.06}_{-0.09}\ \mathrm{km\ s^{-1}}$ (see 
Section~\ref{sect:vsini}). Due to the tidal synchronization, the spin 
period of the visible star should be identical to $P_{\mathrm{orb}}$. With 
such a rapid spin, the visible star is oblate. Hence, the equatorial plane 
$V_{\mathrm{rot}}$ should be evaluated numerically. We adopt \texttt{Phoebe} 
(PHysics Of Eclipsing BinariEs, which is publicly available 
from \url{http://phoebe-project.org}; \cite{Conroy2020}) 
to model $V_{\mathrm{rot}}$ of such a rapid spinning star. To do so, we 
set the \texttt{Phoebe} stellar parameters according to the \texttt{ARIADNE} 
posterior distributions of $M_{\mathrm{vis}}$ and $R_{\mathrm{vis}}$. Then, 
we use \texttt{Phoebe} to calculate the corresponding rotation spread functions 
for a star with a spin period of $0.47992$ days. 
Hence, we can obtain the distribution of the maximal rotation velocity of the 
visible star at the $\Phi=0.75$ phase. From the distribution, we have 
$V_{\mathrm{rot}} = 71\pm 1.8\ \mathrm{km~s^{-1}}$. The corresponding 
inclination angle of J2354 is $i=62^{+3.0}_{-2.6}$ degrees (case I) or 
$i=73^{+5}_{-4}$ degrees (case II). Hence, as shown in Figure~\ref{fig:mass}, 
$M_{\mathrm{inv}}=1.60^{+0.11}_{-0.10}\ M_{\odot}$ (case I) or 
$M_{\mathrm{inv}}=1.40^{+0.09}_{-0.08}\ M_{\odot}$ (case II). 
Some key physical properties of J2354 are summarized in 
Table~\ref{tbl:property}. 

The Roche lobe radius \citep{Paczynski1976} of the visible star is 
$0.462(M_{\mathrm{vis}}/(M_{\mathrm{vis}}+M_{\mathrm{inv}}))^{(1/3)}a=1\ 
R_{\odot}$, where $a=3.2\ R_{\odot}$ is the semi-major axis of the binary. 
The corresponding Roche lobe fill factor is $65\%$, i.e., the Roche lobe 
is far from being filled by the visible star. If there was mass 
transferring from the visible star to the compact object, one would expect 
the Roche lobe filling factor to be as high as $90\%$ \citep{Benvenuto2015}. 
In summary, the birth mass of the unseen compact object is as large as 
$1.4\sim 1.6\ M_{\mathrm{\odot}}$. Given such a large birth mass, the compact 
object in J2354 is probably a neutron star. Although highly unlikely, the compact 
object might also be one of the most massive but cold white dwarfs within $\sim 100$ 
pc \citep{Kilic2021}.

\begin{table*}[h!]
  \centering
  \caption{A summary of the physical properties of J2354. \label{tbl:property}}
  \begin{tabular}{ccc}
  \hline
   & Parameter  & Value \\ 
  \hline
  \textbf{Properties of the source} & &   \\
   Right ascension   &   RA\ [deg]    &     358.736516   \\
   Declination   &   DEC\ [deg]    &    33.940474       \\
   V-band magnitude    &   [mag]    &     13.61$\pm$ 0.02       \\
   Distance    &   $D$\ [pc]     &     127.7$\pm$ 0.3   \\
   Extinction   &   $A_{\rm V}$\ [mag]  &   0.00$^{+0.03}_{-0.00}$  \\
   \hline
   \textbf{Parameters of the visible star} & &  \\
    Mass   &    $M_{\mathrm{vis}}^{g}$\ [$M_{\odot}$]  &   0.73$^{+0.06}_{-0.05}$   \\
    Radius  &  $R_{\mathrm{vis}}$\ [$R_{\odot}$]  &   0.66$^{+0.02}_{-0.01}$  \\
    Surface gravity    &   $\log g$\ [cgs]   &   4.66$\pm$ 0.02  \\
    Effective temperature   &   $T_{\rm eff}$\ [K]  &   4070$^{+30}_{-40}$  \\
    Bolometric luminosity  &   $L_{\mathrm{bol}}$\ [$L_{\odot}$]   &  0.108$\pm$ 0.005   \\
    Projected rotation velocity (case I)  &  $V_{\rm rot}\,\sin i$\ [km s$^{-1}$]   &   63$^{+8}_{-6}$ \\
    Projected rotation velocity (case II)  &  $V_{\rm rot}\,\sin i$\ [km s$^{-1}$]   &  68.37$^{+0.06}_{-0.09}$ \\
    \hline
    \textbf{Parameters of the orbit} & &  \\
    Orbital period   &  $P_{\rm orb}$ \ [days] 	&   0.47992$\pm$ 0.00001  \\
    Eccentricity   &   $e$     &   0.002$\pm$ 0.002  \\
    Center-of-mass $V_{\rm r}$   &   $\gamma$\ [km s$^{-1}$]  &   41$^{+2.4}_{-2.3}$  \\
    $V_{\rm r}$ semi-amplitude of visible star   &  $K_{\mathrm{vis}}$\ [km s$^{-1}$]   &  219.4$\pm$0.5  \\
    $V_{\rm r}$ semi-amplitude of H$\alpha$   &  $K_{\mathrm{H_{\alpha}}}$\ [km s$^{-1}$]   & 216$\pm$2  \\
    Mass function  &   $f(M_{\mathrm{inv}})$\ [$M_{\odot}$]	&  0.525$\pm$ 0.004  \\
    Inclination (case I)  &  $i\ [^{\circ}$]  &  62$^{+3.0}_{-2.6}$ \\
    Inclination (case II)  &  $i\ [^{\circ}$]  &  73$^{+5}_{-4}$ \\
    Minimum mass of the invisible object  & $M_{\mathrm{inv,min}}$\ [$M_{\odot}$]  &   1.29$\pm$ 0.04   \\
    Mass of the invisible object (case I)  & $M_{\mathrm{inv}}$\ [$M_{\odot}$]  &  1.60$^{+0.11}_{-0.10}$  \\
    Mass of the invisible object (case II)  & $M_{\mathrm{inv}}$\ [$M_{\odot}$]  & 1.40$^{+0.09}_{-0.08}$   \\ 
\hline
\end{tabular}
\\
\textbf{Notes.} The reported uncertainties correspond to $1\sigma$ confidence interval. 
\end{table*}

\section{Discussion}
\label{sect:discussion}
\subsection{The Origin of the UV Emission}
\label{sect:swift}
As we showed in Section~\ref{sect:sed-data}, the photometric emission of a 
K7 main sequence star matches the observed optical-to-infrared SED well (see 
Figure~\ref{fig:sed}). However,  the archival GALEX (Galaxy Evolution Explorer) 
NUV fluxes show clear excess beyond the photospheric emission. We argue 
that the NUV excess is produced by the chromospheric activity of the K7 star 
rather than the thermal emission from a white dwarf. 

\begin{figure*}
  \centerline{\includegraphics[keepaspectratio, width=0.5\textwidth]{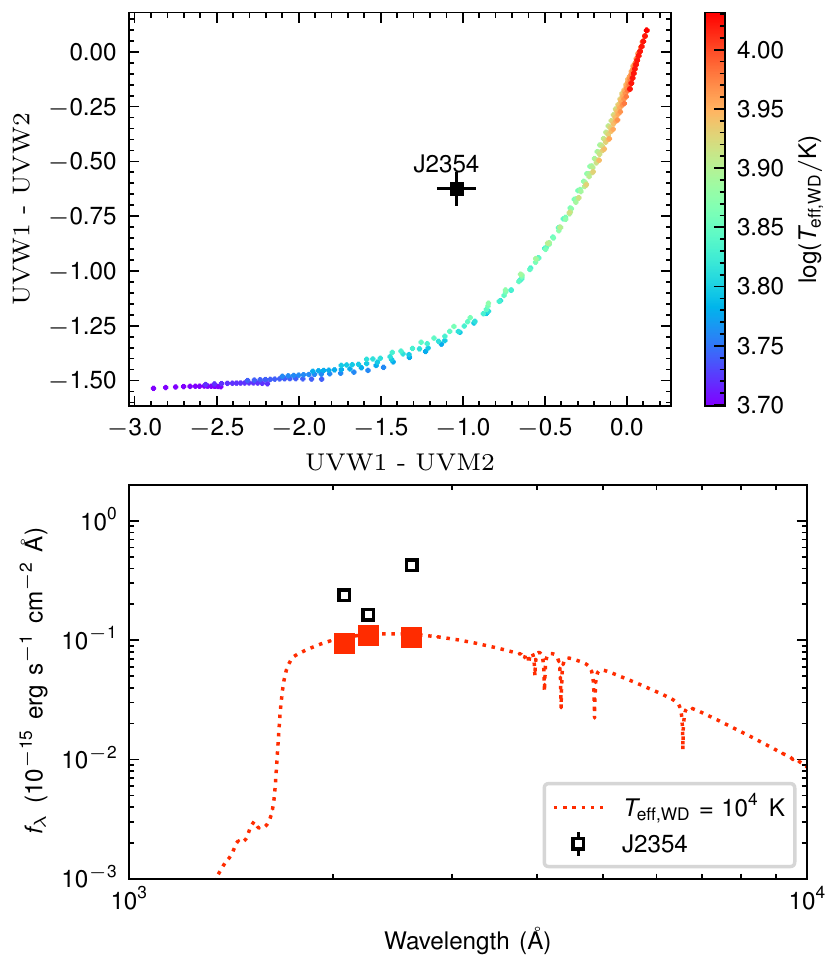}}
  \caption{\textbf{Comparison between the white dwarf templates and the UV excess.} 
  Upper: the UVW1$-$UVM2 colour vs the UVW1$-$UVW2 colour for the templates 
  and J2354. Lower: the WD template with $T_{\mathrm{eff, WD}}= 10,\ 000\ 
  \mathrm{K}$ for a white dwarf with a radius of $0.005\ R_{\odot}$. The filled 
  squares represent the template UVW2, UVW1, and UVM2 magnitudes by convolving 
  the template with the filter response curves. The open squares are for the 
  Swift UVOT observations. Note that the corresponding error bars are too small 
  to be visible. Templates with effective temperatures $> 10,\ 000$ K 
  are inconsistent with the UVOT observations. }
  \label{fig:uv}
\end{figure*}

Six Swift follow-up ToO observations for J2354 are requested and cover various 
orbital phases (Section~\ref{sect:xray-uv}). In these observations, the compact 
object is not eclipsed by the K7 star. The average Swift $UVW1$, $UVM2$, and 
$UVW2$ also show clear excess beyond the photospheric emission. The UV flux 
excess in Swift $UVW1$, $UVM2$, and $UVW2$ and GALEX NUV bands can be obtained 
by subtracting the photosphere SED model from the observed UVW1, UVM2, and 
UVW2 fluxes. We then calculate the colours of the UV excess and compare it 
with the white dwarf templates of \cite{Koester2010} with various temperature. 
To obtain the corresponding UV colours for each white dwarf template, we 
convolve the white dwarf templates of \cite{Koester2010} with the 
UVOT filter response curves to obtain the corresponding UVOT synthetic 
photometries and colours. For all white dwarf templates, the synthetic 
UVOT colours are inconsistent with the observed ones (the left panel of 
Figure~\ref{fig:uv}). We also compare the white dwarf templates of 
\cite{Koester2010} with the SED of the UV excess (the right panel 
of Figure~\ref{fig:uv}). The compact object cannot be a white dwarf that 
is hotter than $10^4$ K; otherwise, the corresponding UV fluxes 
are higher than that observed by Swift. The temperature 
upper limit is crucial for understanding the \texttt{TESS} variability 
and probing the physical nature of the compact object (see 
Section~\ref{sect:variation}). 

The UV excess with a GALEX NUV luminosity of $8\times 10^{29}\ 
\mathrm{erg\ s^{-1}}$ can be well explained by the chromospheric activities 
of the visible star. Indeed, J2354 shows a prominent H$\alpha$ emission 
line whose centre shifts in tandem with the stellar absorption features 
(Figure~\ref{fig:lc}), and H$\alpha$ is a good indicator of chromospheric 
activities \citep{Jones2016, Linsky2017}. The H$\alpha$ luminosities are 
measured by using the LDR8 and LDR9 spectra. To perform the flux 
calibration, the LAMOST spectra are compared with the SED in the 
wavelength range [6520 \AA, 6620 \AA], which covers H$\alpha$. Then, 
the H$\alpha$ flux can be simply measured as $f_{\mathrm{H}\alpha} = 
\int_{\mathrm{wl}}^{\mathrm{wr}} (f(\lambda) - \rm{c}(\lambda)) d\lambda$, 
where wl and wr correspond to the left and right wings of the H$\alpha$ line, 
$f(\lambda)$ is the observed flux, and $\rm{c}(\lambda)$ is the flux of the 
underlying continuum. Note that the continuum flux is obtained by fitting a 
5th-order polynomial to the spectra. The average (over the 22 spectra) 
H$\alpha$ flux is 
$\overline{f}_{\mathrm{H}\alpha} = 2.63\times 10^{-14}\ \rm erg\ s^{-1}\ 
cm^{-2}$. Thus, the H$\alpha$ luminosity is $L_{\mathrm{H\alpha}} = 4\pi D^{2} 
\overline{f}_{\mathrm{H\alpha}} = 5.13\times 10^{28} \ \rm erg\ s^{-1}$, 
where $D=127.7$ pc. Following our previous work \citep{Yi2022}, we find that 
$L_{\mathrm{H\alpha}}/L_{\mathrm{bol}}$ and $L_{\mathrm{NUV}}/L_{\mathrm{bol}}$ 
of J2354 are comparable with isolated M dwarfs \citep{Jones2016}. 

\begin{figure*}
  \centerline{\includegraphics[keepaspectratio, width=0.8\textwidth]{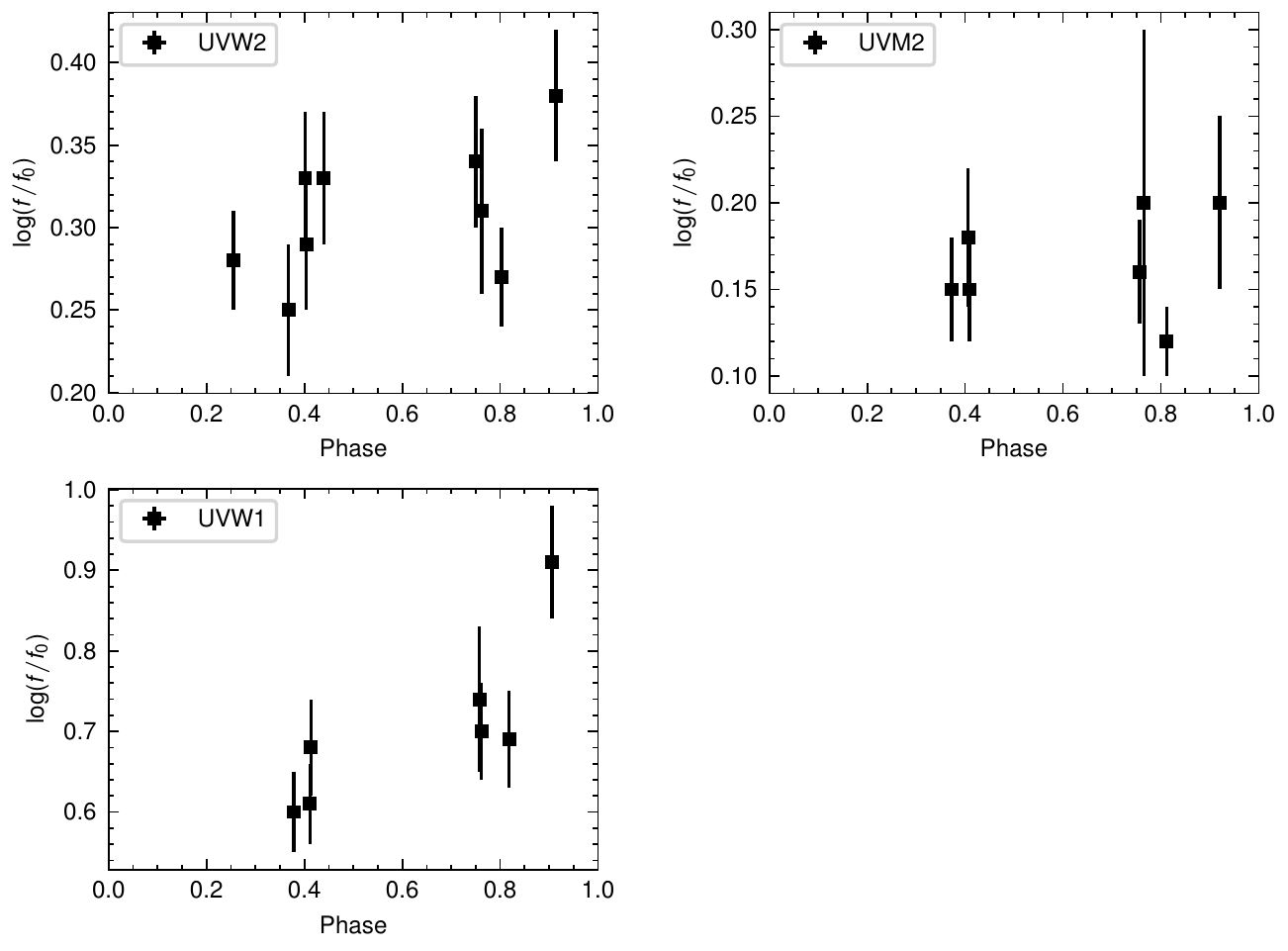}}
  \caption{\textbf{The UV fluxes as a function of the orbital phase of the 
  visible star.} Note that each Swift observation consists of $1\sim 3$ 
  exposures, and the UV fluxes are independently measured in each 
  exposure. Note that $f_0=10^{-15}\ \mathrm{erg\ s^{-1}\ cm^{-2}\ \AA^{-1}}$. 
  In all observations, the compact object is not eclipsed by 
  the visible companion. The UV emission shows evident variability, 
  which is accompanied by an X-ray flare at the orbital phase $\phi\simeq 0.9$. 
  Hence, the UV and X-ray emission is both produced by the visible star. }
  \label{fig:xray}
\end{figure*}

J2354 is undetected by the ROSAT all-sky surveys and lacks Chandra, 
XMM-Newton, or eROSITA public available data. In the first of our 
Swift follow-up observations, for the first time, we detect an 
X-ray counterpart in the optical position of J2354. This X-ray detection 
corresponds to an X-ray flare 
($\log (L_{\mathrm{x,flare}}/[\mathrm{erg\ s^{-1}}])=30.3^{+0.2}_{-0.3}$) 
since the source is undetected in the following second, fourth, and fifth 
observations. In the third observation, the detected X-ray is also much 
less powerful than the first one (Section~\ref{sect:xray-uv}). 
If the X-ray flare is driven by the coronal activity of the visible star, 
one would also expect to detect a UV flare \citep{Tsikoudi2000, Mitra-Kraev2005}. 
Indeed, such a UV flare is observed in the UVOT exposures (Figure~\ref{fig:xray}), 
with the UVM2 luminosity increasing by 
$L_{\mathrm{UVW2,flare}}\sim 3\pm 1.5 \times 10^{29}\ \mathrm{erg\ s^{-1}}$. 
In summary, the X-ray and UV emission are produced by the coronal and 
chromospheric activities of the rapidly spinning visible star 
\citep{Delfosse1998,Stauffer1994}. 

\subsection{Possible origin of the orbital modulation}
\label{sect:variation}

The optical variability of J2354 may offer new clues to the nature of 
the compact object. The model light curve of ellipsoidal modulations 
is calculated for $i=73$ degrees, $M_{\mathrm{vis}}=0.73\ M_{\odot}$, 
and $M_{\mathrm{inv}}=1.4\ M_{\odot}$ via 
\texttt{ELLC} \citep{Maxted2016}; the 
gravity-darkening is fixed to $0.44$ and adopt a quadratic limb-darkening 
law with the following parameters ($0.515$, $0.1754$) for the TESS 
band \citep{Claret2017}. The model 
light curve can be normalized in two ways. That is, its median value 
is the same as the median of the TESS light curve at $\Phi=0.75$ 
(the purple dashed curve in Figure~\ref{fig:lc}; hereafter the minor-peak 
model) or $\Phi=0.25$ (the black dot-dashed curve in Figure~\ref{fig:lc}; 
hereafter the major-peak model). As shown in 
Figure~\ref{fig:lc}, the model light curves can only account for 
the shape around one of the two peaks in the TESS light curve. 

\begin{figure*}
  \centering
  \includegraphics[keepaspectratio, width=0.5\textwidth]{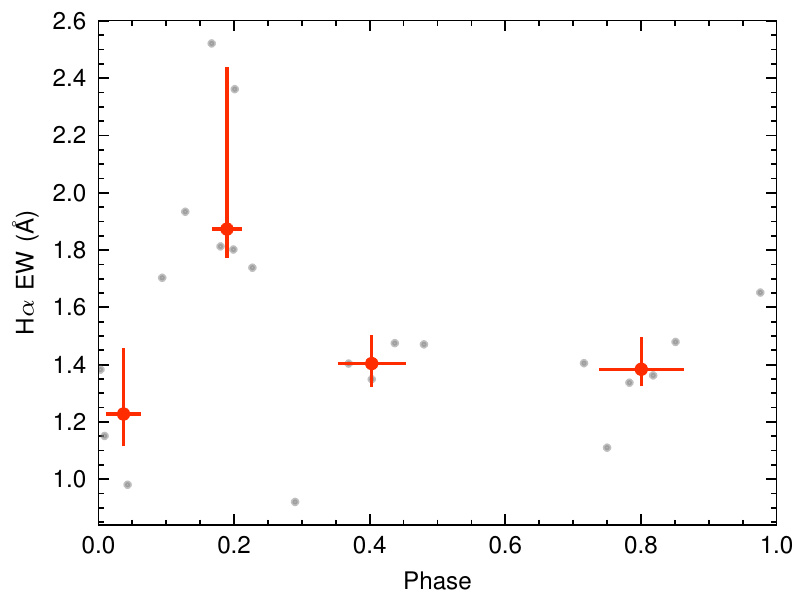}
  \caption{\textbf{The Equivalent Width (EW) of the H$\alpha$ emission 
  line at various phases.} The gray dots represent the EW in each 
  observation; the red circles with errorbars correspond to the 
  averaged EWs and their $1\sigma$ uncertainties. The EW has a 
  peak around $\Phi\simeq 0.25$.}
  \label{fig:ha_ew}
\end{figure*}

To explain the full TESS light curve, non-uniform temperature 
distributions over the stellar surface should be introduced. 
For instance, the major-peak model with a giant starspot that is 
observable from $\Phi=0.4$ to $\Phi=1$ can resolve the discrepancy 
between the major-peak model and the TESS data. At these phases, 
strong magnetic fields in the starspots can induce strong photospheric 
activities. Hence, the $H\alpha$ EWs at these phases (from $\Phi=0.4$ 
to $\Phi=1$) are expected to be larger than those at other phases. This 
expectation is in contrast with observations (Figure~\ref{fig:ha_ew}). 
Alternatively, the minor-peak model with the additional heating from 
the companion also has the potential to explain the full TESS 
light curve. Indeed, the asymmetric TESS phase-folded light 
curve of J2354 are often observed 
in redback millisecond pulsars (e.g., PSR 3FGL J2039.6-5618 
\cite{Strader2019} and 3FGL J0212.1+5320 \cite{Shahbaz2017}) or 
detached magnetic white dwarfs \citep{Parsons2021}. 
The additional heating can be intra-binary shocks 
\citep[e.g.,][]{Romani2016} and is responsible for the major flux 
peak at $\Phi=0.25$. This speculation is further supported by 
the fact that the equivalent width (EW) of the $H\alpha$ emission line 
is strongest around $\Phi=0.25$. In summary, 
we find tentative evidence that the orbital modulations are driven 
by the additional heating, e.g., intra-binary shocks. A detailed 
comparison between the intra-binary shock model and the TESS light 
curves can constrain the shock properties \citep[see, e.g., Figure 4 
in][]{Romani2016}, but is beyond the scope of this 
work. 

\begin{figure*}
  \centering
  \includegraphics[keepaspectratio, width=0.5\textwidth]{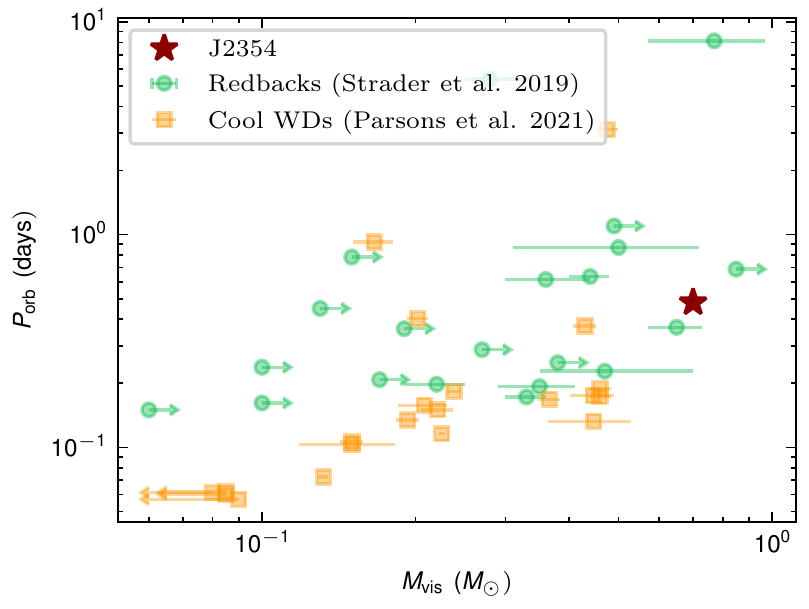}\\
  \caption{\textbf{The orbital period vs the mass of the visible star.} The orange 
  squares are for cool but massive WDs in \cite{Parsons2021}; the green 
  circles represent the redbacks in \cite{Strader2019}; symbols with arrows 
  correspond to the $1\sigma$ uncertainties. The mass of 
  the visible star in J2354 is similar to redbacks rather than cool and 
  massive WDs. }
  \label{fig:companion}
\end{figure*}

The compact object in J2354 can only be a neutron star rather than 
a cold white dwarf if there is indeed additional heating. A cold 
white dwarf with $T_{\mathrm{eff, WD}}\lesssim 10^4$ K cannot 
drive strong winds; its thermal radiation to the surface of the 
visible star is about $\sigma_{\mathrm{SB}}T_{\mathrm{eff, WD}}^4
(R_{\mathrm{WD}}/a_{\mathrm{orb}})^2(4\pi R_{\mathrm{vis}}^2) 
\lesssim 10^{-4} L_{\odot}$, which is clearly too small to heat 
up the visible star. Instead, a neutron star can easily have 
a powerful and variable pulsar wind to significantly heat the 
visible star \citep[e.g.,][]{Cho2018}. In addition, J2354's 
$M_{\mathrm{inv}}$ and $P_{\mathrm{orb}}$ are similar to several 
redbacks in \cite{Strader2019}. In contrast, none of cool 
and massive WDs in \cite{Parsons2021} hosts a visible star as 
massive as that in J2354 (Figure~\ref{fig:companion}). 

\subsection{Searching for radio and $\gamma$-ray counterparts}
\label{sect:fast}

We performed a total of 1.7 hours of targeted exceptionally sensitive radio 
follow-up observation for the pulsar search using FAST in the following three 
sessions: (1) 5 Sep 2021 17:30:00 to 18:30:00 UTC (Coordinated Universal Time); 
(2) 30 Aug 2022 17:55:00 to 18:15:00 UTC; (3) 25 Sep 2022 16:20:00 to 16:40:00 
UTC. The orbital-phase ranges of the visible star in these sessions are [$0.20$, 
$0.28$], [$0.20$, $0.22$], and [$0.24$, $0.26$]. In these phases, the unseen 
compact object is not ``shielded'' by the K7 star. The first one minute and the 
last one minute were the calibration signal injection time for the flux and 
polarization calibration in each of the observations. Observations taken at FAST 
are using the centre beam of the 19-beams L-band receiver, the frequency range 
is $1.05$--$1.45$ GHz with the average system temperature 25 K \citep{Jiang2020}. 
Observational data are recorded in pulsar search mode, stored in PSRFITS format 
\citep{Hotan2004}. We performed two types of data processing during the 
observational campaign: 

I. Dedicated and blind search:

Based on the Galactic electron density model of NE2001 \citep{Cordes2002} and 
YMW16 \citep{Yao2017}, we estimate, for the distance $D$=127.7 $\pm$ 0.3 pc, 
the corresponding with a dispersion measure (DM) of $\sim$ 1.4 pc cm$^{-3}$, 
and the line of sight maximal Galactic $\mathrm{DM_{max}}=50\ \mathrm{pc\ cm^{-3}}$. 
Meanwhile, the X-ray column density is fixed at $\mathrm{N_{H}}=5\times 10^{20}\ 
\mathrm{cm^{-2}}$ corresponding with $\mathrm{DM}\simeq 20\ \mathrm{pc\ cm^{-3}}$ 
estimated by the empirical linear relation \citep{He2013}. Due to the model 
dependence and for the sake of robustness, we created de-dispersed time series 
for each pseudo-pointing over a range of DMs, from $0$--$300\ \mathrm{pc\ cm^{-3}}$, 
which is a factor of six larger than $\mathrm{DM_{max}}$. For each of the trial 
DMs, we searched for a periodical signal and first two order acceleration in the 
power spectrum based on the PRESTO \citep{Ransom2003} pipeline \citep{Wang2021}. 
We checked all the pulsar candidates of signal-to-noise ratios (SNR) $>$ 6 and 
removed the narrowband radio frequency interferences (RFIs). 

\begin{figure*}
  \centering
  \includegraphics[keepaspectratio, width=0.5\textwidth]{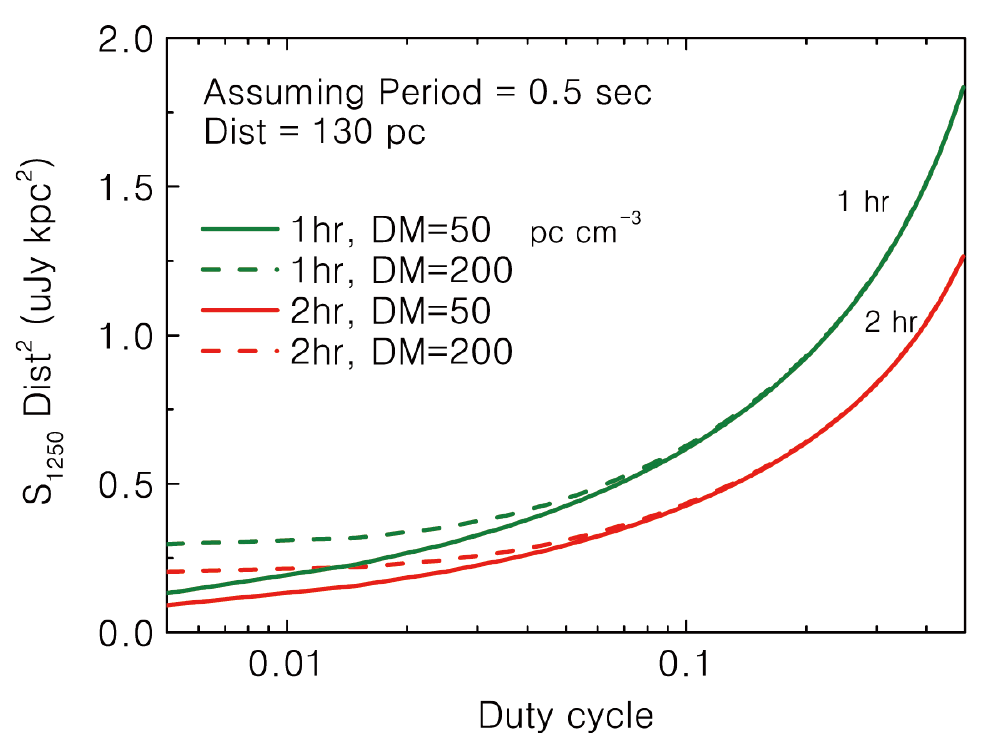} \\
  \caption{\textbf{The sensitivity limits of 6$\sigma$ detection at 1250MHz vs. pulse duty 
  cycle (assuming a young pulsar rotation period $\sim$0.5 second).} The green and 
  red curves are for the upper limit of flux densities with one-hour and two-hour 
  integration time, respectively. The solid and dashed lines represent the 
  sensitivity for DM = 50 pc cm$^{-3}$ and 200 pc cm$^{-3}$, respectively. 
  For a nearby pulsar, the effective width 
  $W_{\mathrm{eff}}\sim \sqrt{W^2_{\mathrm{int}}+\tau^2_{\mathrm{smear}}}$, and 
  the inter-channel DM smearing time could be estimated by 
  $\tau_{\mathrm{smear}}=8.3\mathrm{DM}(\delta f/\mathrm{MHz})/(f/\mathrm{GHz})^3\ 
  \mu s$, where $\delta f=0.122\ \mathrm{MHz}$ is the frequency channel width. The 
  detected pulse profile will be broadened due to that the inter-channel DM smearing, 
  thus affecting the duty cycle. As a result, the sensitivity limit depends weakly 
  upon DM, if a small duty cycle is assumed.}
  \label{fig:fast}
\end{figure*}

Both the periodical radio pulsations and single-pulse blind searches were performed 
for each observing epoch, but resulted in non-detections for all sessions. In 
Figure~\ref{fig:fast}, we estimated the FAST detected sensitivity dependence of DMs 
and pulse duty cycle based on the radiometer equation. Furthermore, we calibrated the 
noise level of the baseline, and then measured the amount of pulsed flux above the 
baseline, giving the 6$\sigma$ upper limit of flux density measurement of 
25 $\pm$ 5 $\mu$Jy in session (1) for persistent radio pulsations (assuming a pulse 
duty cycle of $0.05$--$0.3$). The time interval between sessions (2) and (3) is nearly 
one month, and the effect of interstellar scintillation can be well excluded; the 
upper limit of the flux density can be given for the average of the two measurements 
of 12.5 $\pm$ 2.0 $\mu$Jy in session (2-3). 

II. Single pulse search:

The above search strategy was continued to be used to de-disperse the time series 
for single pulse search and flux calibration. We used PRESTO and HEIMDALL 
\citep[\url{https://sourceforge.net/p/heimdall-astro/wiki/Home/};][]{Champion2016} 
software. A zero-DM matched filter was applied to mitigate RFI in the blind search. 
All possible candidates were plotted, then be confirmed as RFIs by manual check. No 
pulsed radio emission with a dispersive signature was detected with an SNR $> 6$. The 
upper limit of pulsed radio emission is $\sim 0.015$ Jy ms assuming a 1 ms wide burst 
in terms of integrated flux (fluence). 

In summary, radio pulsars or persistent radio emission are not detected in the FAST 
data. The $6\sigma$ upper limit of the potential pulse power at $1.4$ GHz is 
therefore $<3.4\times 10^{23}\ \mathrm{erg\ s^{-1}}$. Unlike redbacks, 
J2354 also lacks $\gamma$-ray counterparts in the Fermi's 4FGL catalogue 
\citep{Abdollahi2020}. Hence, the candidate is likely to be a 
non-beaming neutron star. Such sources will be easily missed in radio or $\gamma$-ray 
observations and can only be unearthed by optical time-domain surveys. 

\subsection{The Trajectory of J2354.} 
\label{sect:movement}
\begin{figure*}
    \centerline{\includegraphics[keepaspectratio, width=0.8\textwidth]{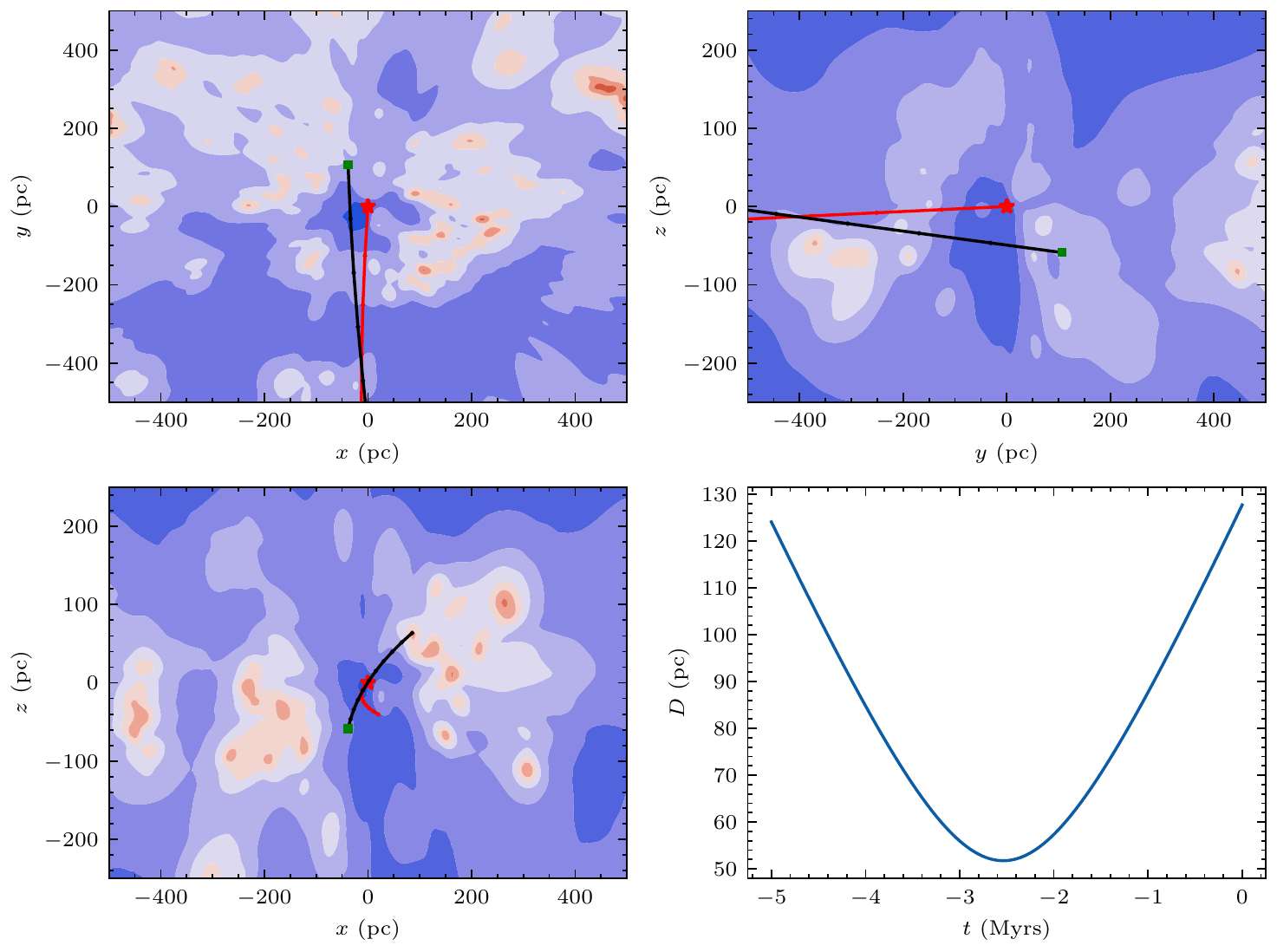}}
    \caption{\textbf{The current position (green squares) and the trajectory of 
    J2354 in the past five Myrs (black dots and curve).} The upper-left, 
    upper-right, and lower-left 
    are for the $xy$-plane, $yz$-plane, and $xz$-plane, respectively. The red 
    star and curve indicate the current and historical positions of our Sun. The 
    filled contours represent 
    the ``Stilism'' tridimensional map of the local ISM \citep{Capitanio2017}. 
    The regions with low-density ISM are indicated as deep blue colours. The 
    Local Bubble, a low-density cavity around our Sun, is evident in the figure. 
    The lower-right panel shows the historical distance between J2354 and our Sun. }
    \label{fig:bubble}
\end{figure*}

The distribution of the local ISM shows a bubble structure (i.e., the Local 
Bubble). The tridimensional map of the local ISM can be constructed from the 
inversion of the dust and gas absorption via the STructuring by Inversion the 
Local Interstellar Medium (hereafter Stilism; \url{https://stilism.obspm.fr/}) 
project \citep{Capitanio2017}. In this map, the position of our Sun is 
$x, y, z=0$ pc, where $z$ is perpendicular to the galactic plane. The $xy$-plane 
(i.e., the galactic plane with $z=0$ pc), $xz$-plane (with $y=0$ pc), and $yz$-plane 
(with $x=0$ pc) cuts of the ``Stilism'' map are shown in Figure~\ref{fig:bubble}. 
It is evident that our Solar system and J2354 reside in the Local Bubble 
(Figure~\ref{fig:bubble}). The bubble is speculated to be created by multiple nearby 
supernovae \citep{Smith2001,Schulreich2017}. The possible remnants, neutron 
stars or black holes, of these supernovae remain hidden. 

With the \textit{Gaia} DR3 
\citep{Gaia} proper-motion data ($23.21\pm 0.02\ \mathrm{mas\ yr^{-1}}$ and 
$-18.93\pm 0.01\ \mathrm{mas\ yr^{-1}}$ in the RA and DEC directions) and our 
best-fitting systematic radial-velocity (i.e., $\gamma = 41\ \mathrm{km\ s^{-1}}$) 
for J2354, the trajectory of J2354 in our Galaxy in the past five Myrs can be 
calculated by considering the influence of the Milky-Way potential. The Milky 
Way potential consists of four components, i.e., a spherical nucleus, a bulge, 
a disk model \citep{Bovy2015} and a Navarro-Frenk-White halo. The trajectory 
is integrated backwards with a time step of $0.05$ Myrs via the \textit{Gala} 
\citep{Gala} code. We also calculate the three Galactic space velocities, i.e., 
$U$ (the velocity towards 
the Galactic centre), $V$ (the velocity along the Galactic rotation), and $W$ 
(the velocity in the direction of the North Galactic Pole). Then, we find that, 
according to the $UVW$ velocities \citep{Li2017}, J2354 belongs to the Galactic 
thin disk. 

The historical distances between J2354 and our Sun in the past $5$ 
Myrs are closer than the present distance, and the minimum historical distance 
is only $52$ pc (i.e., two times smaller than its present distance). Hence, the 
neutron star candidate in J2354 is one of the nearest neutron stars in binaries. 
Then, we calculate the trajectory of J2354 in the three cuts of the ``Stilism'' 
map. In the past five Myrs, J2354 has passed through the Local Bubble. The 
detection of J2354 demonstrates the possibility of improving the demographics 
of near-Earth supernovae in the era of time-domain astronomy, which is vital 
for our understanding of the chemical enrichment history and the energy and 
gas recycling around the Solar system. 

\subsection{Alternative Scenario}\label{sect:wd-model}

We cannot entirely exclude the possibility that J2354 harbours an ultramassive 
but cold white dwarf. If so, the candidate 
in J2354 is one of the most massive nearby ($\lesssim 100$ pc) white dwarfs 
\citep{Kilic2021}. The other scenario involves a hierarchical triple system, i.e., 
the visible star orbits around two unseen $\sim 0.7\ M_{\odot}$ white dwarfs. 
According to the observed short orbital period and large 
mass function, the semi-major axis of the visible star's orbit is only 
$a= 3.2\ R_{\odot}$. For such a triple system to be stable, the 
semi-major axis of the orbit of the two unseen white dwarfs should be much smaller 
than the distance between the white dwarfs and the barycentre of J2354, which 
is $\sim 0.7/(0.7+1.4)a=1.1\ R_{\odot}$. The corresponding orbital period of the 
two white dwarfs is $\lesssim 10^4$ seconds. Then, the system 
is a promising gravitational wave source for space-based gravitational wave 
observatories \citep{Littenberg2019, Taiji, Tianqin}. We 
stress that, in the hierarchical triple system scenario, the two white dwarfs 
should also be cold ($\lesssim 10^4$ K). Hence, the additional heating from 
the white dwarfs is again too weak to explain the observed TESS light 
curve (see Section~\ref{sect:variation}).

\section{Summary}
\label{sect:sum}

In this work, we have provided evidence that J2354 likely hosts a neutron star 
with a distance to Earth of $127.7\pm 0.3\ \mathrm{pc}$. Our results can be 
summarized as follows. 
\begin{itemize}
  \item We have used the doppler spectroscopy to measure the mass function of 
  the unseen neutron star (Sections~\ref{sect:lamost} \& \ref{sect:fmass}), i.e., 
  $f(M_{\rm{inv}})=0.525\pm 0.004\ M_{\odot}$. 
  \item We have determined the inclination angle of J2354 by measuring the rotational 
  broadening velocity in the high-resolution CFHT spectra (Section~\ref{sect:cfht}). 
  As a result, the mass of the neutron star candidate, 
  $M_{\mathrm{inv}}=1.4\sim 1.6 M_{\odot}$ (Section~\ref{sect:minv}). 
  \item We have, for the first time, detected the X-ray counterpart of J2354 via 
  \texttt{Swift}. Our joint analyses of the \texttt{Swift} X-ray and UVOT 
  variability suggest that the X-ray and UV emission are produced by the 
  coronal and chromospheric activities of the K7 star (Section~\ref{sect:swift}). 
  \item Pulsating or persistent radio emission cannot be detected from the 
  $1.7$-hour exceptionally sensitive FAST observations at the $6\sigma$ 
  flux upper limit of $12.54\ \mathrm{\mu Jy}$ (Section~\ref{sect:fast}). 
  \item We have used the \texttt{Gaia} proper-motion measurements to calculate 
  the historical positions of J2354 in the past five Myrs. It is evident that 
  J2354 passes through the Local Bubble. The minimum distance between the Sun 
  and J2354 is only $52$ pc (Section~\ref{sect:movement}). 
\end{itemize}

\footnotesize{\textbf{Acknowledgements.}}
We thank Hong-Gang Yang, Yi-Han Song, Xuan Fang, Shuai Liu, and 
Yi-Ze Dong for beneficial discussions. 
W.M.G. acknowledges support from the National Key R\&D Program of 
China under grant 2021YFA1600401, and the National Natural Science Foundation of China 
(NSFC) under grants 11925301 and 12033006. M.Y.S. acknowledges support from the NSFC 
under grants 11973002 and 12322303. Z.X.Z. acknowledges support from the NSFC under 
grant 12103041. J.F.L. acknowledges support from the NSFC under grants 11988101 and 
11933004. P.W. acknowledges support from the NSFC under grant U2031117. J.F.Wang 
acknowledges support from the NSFC under grant 12033004. J.F.Wu acknowledges 
support from the NSFC under grant 12273029. W.M.G., J.F.Wang, and J.F.Wu acknowledge 
support from the NSFC under grant 12221003. J.Z. acknowledges support from the NSFC 
under grant 11933008. J.R.S. acknowledges support from the NSFC under grant 12090044. 
X.D.L. acknowledges support from the NSFC under grants 12041301 and 12121003.

This work uses the LAMOST spectra. Guoshoujing Telescope (the Large Sky 
Area Multi-Object Fiber Spectroscopic Telescope, LAMOST) is a National Major 
Scientific Project built by the Chinese Academy of Sciences. Funding for the 
project has been provided by the National Development and Reform Commission. 
LAMOST is operated and managed by the National Astronomical Observatories, 
Chinese Academy of Sciences. This work includes public data collected by the 
TESS mission, the NASA Galaxy Evolution Explorer (GALEX), 
SDSS, the Pan-STARRS1 Surveys (PS1), the Two Micron All Sky Survey(2MASS), the 
Wide-field Infrared Survey Explorer (WISE), the Zwicky Transient Facility 
(ZTF) project, ASAS-SN, the European Space Agency (ESA) mission Gaia, TESS, and 
the APASS database. This work uses data obtained through the Telescope Access Program 
(TAP), which has been funded by the TAP member institutes. This work made use of the 
data from FAST. FAST is a Chinese national mega-science facility, operated by National 
Astronomical Observatories, Chinese Academy of Sciences. This work uses data from 
Swift Target of Opportunity observations.  

\footnotesize{\textbf{Conflict of interest.}}
The authors declare that they have no conflict of interest.

\bibliographystyle{scpma}
\bibliography{ms.bib}

\end{multicols}

\end{document}